\documentclass[format=acmsmall, review=false, screen=true]{acmart}

\usepackage{paralist}
\usepackage{multirow}
\usepackage{booktabs}
\usepackage{color}
\usepackage{threeparttable}
\usepackage{diagbox}
\usepackage{makecell}
\usepackage{pifont}
\usepackage{tikz-qtree}
\usepackage{enumitem}
\usetikzlibrary{positioning}

\setcopyright{acmcopyright}
\acmJournal{CSUR}
\acmYear{2018} \acmVolume{1} \acmNumber{1} \acmArticle{1} \acmMonth{1} 
\acmPrice{\$15.00}\acmDOI{10.1145/3204947}

\received{May 2016}
\received[revised]{October 2017}
\received[accepted]{April 2018}

\begin{document}

\title[Analytics for the Internet of Things: A Survey]{Analytics for the Internet of Things: A Survey}  

\author{Eugene Siow}
\affiliation{%
  \institution{University of Southampton}
  \country{UK}}
\email{eugene.siow@soton.ac.uk}
\author{Thanassis Tiropanis}
\affiliation{%
  \institution{University of Southampton}
  \country{UK}}
\author{Wendy Hall}
\affiliation{%
  \institution{University of Southampton}
  \country{UK}}

\begin{abstract}
The Internet of Things (IoT) envisions a world-wide, interconnected network of smart physical entities. These physical entities generate a large amount of data in operation and as the IoT gains momentum in terms of deployment, the combined scale of those data seems destined to continue to grow. Increasingly, applications for the IoT involve analytics. Data analytics is the process of deriving knowledge from data, generating value like actionable insights from them. This article reviews work in the IoT and big data analytics from the perspective of their utility in creating efficient, effective and innovative applications and services for a wide spectrum of domains. We review the broad vision for the IoT as it is shaped in various communities, examine the application of data analytics across IoT domains, provide a categorisation of analytic approaches and propose a layered taxonomy from IoT data to analytics. This taxonomy provides us with insights on the appropriateness of analytical techniques, which in turn shapes a survey of enabling technology and infrastructure for IoT analytics. Finally, we look at some tradeoffs for analytics in the IoT that can shape future research.
\end{abstract}

%
%


%
%





\maketitle

\renewcommand{\shortauthors}{E. Siow et al.}

\newcommand{\quotes}[1]{``#1''}

\section{Introduction}
\label{sec:intro}

\begin{figure}[!h]
	\includegraphics[width=0.8\textwidth]{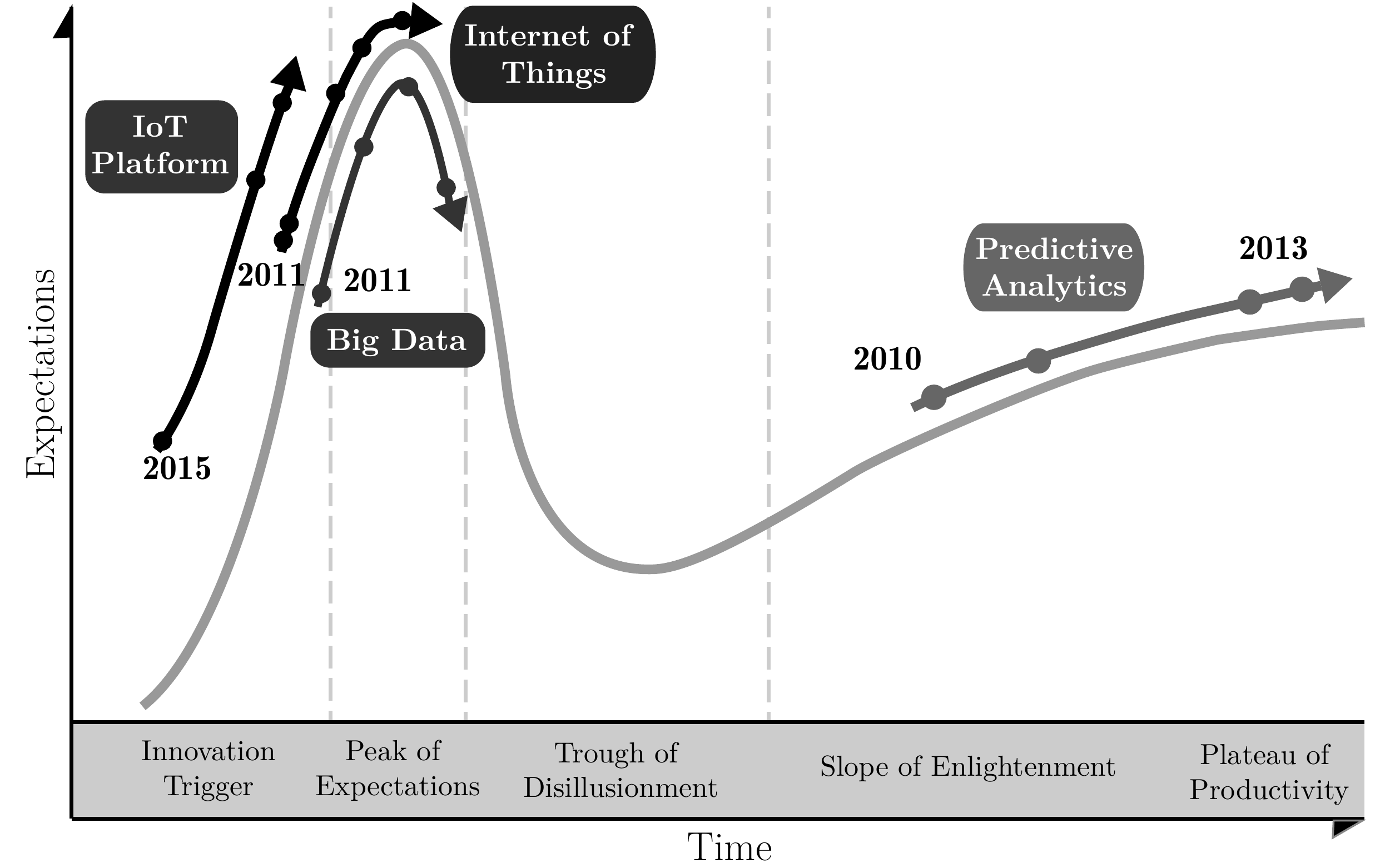}
	\caption{Aggregated Gartner Hype Cycle of Technologies from 2010 \cite{Pettey2010,Pettey2011,Pettey2012,Rivera2013,Rivera2014,Rivera2015,Forni2016,Panetta2017}}
	\label{fig:hypecycle}
\end{figure}

The Internet of Things (IoT) has been gaining momentum in both the industry and research communities due to an explosion in the number of smart mobile devices and sensors and the potential applications of the data produced from a wide spectrum of domains. In their 2013 report, McKinsey note a 300\% growth in connected IoT devices in the last five years and rate the potential economic impact of the IoT at \$2.7 trillion to \$6.2 trillion annually by 2025 \cite{Manyika2013}. These figures grew to \$4 trillion and \$11 trillion in 2015 \cite{Manyika2015}. A study of Gartner's 2010 to 2017 hype cycle reports, which we aggregate in Fig.~\ref{fig:hypecycle}, shows the advent of the IoT, steady growth, expansion and creation of new technology areas like the IoT platform. Another interesting technology that exceeds the IoT in momentum on the hype cycle is that of big data, which the IoT serves as a source and sink of.

Big data is data that are \textit{too big} (volume), \textit{too fast} (velocity) and \textit{too diverse} (variety) \cite{Madden2012}. In the context of the IoT, we see an example of volume in the DEBS 2014 Grand Challenge \cite{Ag2014}, where data from 40 houses with smart plugs produced 4 billion events in a month \cite{Fernandez2014}, given that a 2011 census showed that there were 26.4 million households in the United Kingdom \cite{OfficeofNationalStatistics2013}, the projected data size of 2.64 quadrillion (short scale) per month if every house had a meter, is a good example of \textit{too big} data. In the IoT use cases of intelligent transportation systems \cite{VanNunen2008,OHara2012} and telecommunication, data streams can come in \textit{too fast} for processing, representing a data \textit{velocity} problem. Finally, \textit{too diverse} is the catchall term used to describe the presence of heterogenous data sources in the IoT that make it difficult for existing tools to analyse them. In a 2014 survey of data scientists, 71\% interviewed said that analytics is becoming increasingly difficult due to the variety and types of data sources \cite{Paradigm42014}. An example is in the personal health care use case of the IoT \cite{Niewolny2013}, where unstructured textual electronic health records, connected mobile devices and sensors \cite{Amendola2014} all add to the variety problem.

Analytics is the science or method of using analysis to examine something complex \cite{Dictionarya}. When applied to data, analytics is the process of deriving (the analysis step) knowledge and insights from data (something complex). The evolution to the concept of analytics we see today can be traced back to 1962. Tukey first defined data \textit{analysis} as procedures for analysing data, techniques for interpreting the results, data gathering that makes analysis easier, more precise and accurate and finally, all the related machinery and statistical methods used \cite{Tukey1962}. In 1996, Fayyad \emph{et~al.} published an article explaining Knowledge Discovery in Databases (KDD) as \quotes{the overall process of discovering useful knowledge from data} where data mining serves aa a step in this process - \quotes{the application of specific algorithms for extracting patterns from data} \cite{Fayyad1996}. In 2006, Davenport introduced analytics as quantitative, statistical or predictive models to analyse business problems like financial performance or supply chains and stressed its emergence as a fact-based decision-making tool in businesses \cite{Davenport2006}. In 2009, Varian highlighted the ability to take data and \quotes{understand it, process it, extract value from it, visualise it and communicate it}, as a hugely important skill in the coming decade \cite{Varian2009}. In 2013, Davenport  introduced the concepts of Analytics 1.0, traditional analytics, 2.0, the development of big data technology and 3.0 where this big data technology is integrated agilely with analytics, yielding rapid insights and business impact \cite{Davenport2013}. 

To better understand each of these areas, the IoT, Big Data and Analytics, and their intersection, we look chronologically at the existing reviews and surveys on these topics. This will help to establish the need for our review from the new dimension of analytics on the IoT especially in big data scenarios. A summary of the reviews is shown in Table \ref{table:surveys}.

\begin{table}[th]%
\caption{Chronological Summary of Previous Surveys in the IoT, Big Data and Analytics}
\label{table:surveys}
\newcommand*\OK{\ding{51}}
\setlength{\tabcolsep}{8pt}
\begin{minipage}{\columnwidth}
\begin{center}
\begin{tabular}{llcccl}
  \toprule
Year 	& 	Reference 									& 	$I$ 	& 	$B$ 	& 	$A$ 	& Description\\
\midrule
2010 	& Atzori \emph{et~al.} \cite{Atzori2010} 		& \OK 	& 		& 		& Vision, Apps\\
 		& Sharma \emph{et~al.} \cite{Sharma2010} 		& 		& 		& \OK 	& Business Analytics \\
2012 	& Miorandi \emph{et~al.} \cite{Miorandi2012} 	& \OK 	& 		& 		& Vision, Challenges\\
 		& Barnaghi \emph{et~al.} \cite{Barnaghi2012} 	& \OK 	& 		& 		& Semantics\\
 		& Chen \emph{et~al.} \cite{Chen2012} 			& 		& 		& \OK 	& Business Analytics\\
2013 	& Sagiroglu \emph{et~al.} \cite{Sagiroglu2013} 	&  		& \OK 	& 		& Problems, Techniques \\
 		& Vermesan \emph{et~al.} \cite{OvidiuVermesan2013} & \OK 	& 		& 	& Vision, Apps, Governance\\
2014 	& Perera \emph{et~al.} \cite{Perera2014} 		&  		& \OK 	& 		& Context-aware\\
 		& Zanella \emph{et~al.} \cite{Zanella2014} 		& \OK 	& 		& 		& Smart Cities\\
 		& Xu \emph{et~al.} \cite{Xu2014} 				& \OK 	& 		& 		& Industries\\
 		& Zhou \emph{et~al.} \cite{Zhou2014}				&  		& \OK 	& \OK 	& Big Data Analytics Challenges\\
 		& Kambatla \emph{et~al.} \cite{Kambatla2014}		&		& \OK 	& \OK 	& Big Data Analytics Trends\\
 		& Chen \emph{et~al.} \cite{Chen2014a} 			& \OK 	& \OK 	& \OK 	& Big Data Analytics\\
 		& Stankovic \cite{Stankovic2014}					& \OK 	& 		& 		& Directions\\
2015	& Al-Fuqaha \emph{et~al.} \cite{Al-Fuqaha2015}	& \OK	& \OK	&		& Protocols,Challenges,Apps\\
		& Granjal \emph{et~al.} \cite{Granjal2015}		& \OK	& 		&		& Security Protocols, Challenges\\
2016	& Ray \cite{Ray2016}							& \OK	& 		&\OK	& IoT Architectures\\
		& Razzaque  \emph{et~al.} \cite{Razzaque2016}	& \OK	& 		& 		& Middleware for IoT\\
2017 	& Akoka \emph{et~al.} \cite{Akoka2017}			& 		& \OK	&		& Big Data Research Trends\\
		& Lin \emph{et~al.} \cite{Lin2017}				& \OK	& 		& 		& Fog Computing Architecture\\	
		& Farahzadia \emph{et~al.} \cite{Farahzadia2017}	& \OK	& \OK	& 		& Middleware for Cloud IoT\\
		& Sethi \emph{et~al.} \cite{Sethi2017}			& \OK	& 	 	& 		& IoT Architectures, Apps\\
\bottomrule
\end{tabular}
\end{center}
\centering
\footnotesize{\textbf{Legend:} $I$=IoT, $B$=Big Data, $A$=Analytics}
\end{minipage}
\end{table}%

In 2010, Atzori \emph{et~al.} \cite{Atzori2010} survey the vision of the IoT, the enabling technologies and potential applications while identifying three perspectives: Things, Semantics and Network. Sharma \emph{et~al.} \cite{Sharma2010} study analytics applications in the industry and propose a framework of how business analytics can be applied to processes for organisations to gain a sustainable, competitive advantage. 

In 2012, Miorandi \emph{et~al.} \cite{Miorandi2012} survey the IoT mainly from the perspective of the key issues and research challenges and some initiatives going on to address them. Barnaghi \emph{et~al.} \cite{Barnaghi2012} look at developments in the semantic web community, analysing the advantages of semantics but also highlighting the challenges they face and review work on applying semantics to the IoT. Chen \emph{et~al.} \cite{Chen2012} study, using bibliometrics, some of the key research areas in business intelligence and analytics, some application areas and propose a framework to classify them.

In 2013, Sagiroglu \emph{et~al.} \cite{Sagiroglu2013} give an overview of the big data problem, methods to handle the big data, analysis techniques and challenges. Vermesan \emph{et~al.} \cite{OvidiuVermesan2013} look at the vision, applications, governance and challenges of the IoT and some proposed solutions like semantics. 

In 2014, Perera \emph{et~al.} \cite{Perera2014} present a study of context-aware computing and discuss how it can be applied to the IoT. Zanella \emph{et~al.} \cite{Zanella2014} survey the enabling infrastructure and architecture for the Internet of Things in an urban, connected, smart city scenario while Xu \emph{et~al.} \cite{Xu2014} review the development of IoT technologies for industries. Zhou \emph{et~al.} \cite{Zhou2014} discuss the challenges brought to data analytics by big data from the perspective of various applications while Kambatla \emph{et~al.} \cite{Kambatla2014} discuss trends with a focus on hardware and software platforms, virtualisation and application scopes for analytics. Another big data survey is done by Chen \emph{et~al.} \cite{Chen2014a} who look at challenges and work done from each stage of \quotes{data generation, data acquisition, data storage, and data analysis}. They also look at applications of big data briefly, where one such area is the IoT. Finally, Stankovic \cite{Stankovic2014} proposes a set of research directions and considerations for future research on the IoT.

In 2015, Granjal \emph{et~al.} \cite{Granjal2015} survey existing protocols for protecting communications on the IoT, comparing against a set of fundamental security requirements, and highlight the open challenges and strategies for future research work. Al-Fuqaha \emph{et~al.} \cite{Al-Fuqaha2015} focus on giving a thorough summary of protocols for the IoT and how they work together for applications in big data scenarios.

In 2016, Ray \cite{Ray2016} surveys domain-specific architectures for the IoT providing a brief summary of whether cloud platforms in the IoT support data analytics. Razzaque \emph{et~al.} \cite{Razzaque2016} survey middleware platforms for the IoT against a set of comprehensive service and architectural requirements.

In 2017, Akoka \emph{et~al.} \cite{Akoka2017} perform a systematic mapping study, a method for structuring a research field, to classify big data academic research and identify trends in the research. Both analytics and the IoT were identified as popular topics. Reviews by Lin \emph{et~al.} \cite{Lin2017} and Farahzadia \emph{et~al.} \cite{Farahzadia2017} focus on specific IoT research areas of fog computing architectures and middleware for cloud computing platforms. Sethi \emph{et~al.} \cite{Sethi2017} take the approach of surveying IoT architectures, protocols and applications which help them organise a taxonomy of IoT research.

One can see that the vision of the IoT through these surveys is still very much about interconnecting physical objects  with protocols, however, the introduction of Big Data and Analytics has meant that there has been a broadening of focus from communications technologies to applications with impact, scalability and utilising context within cross-domain use cases like the smart city while also coalescing around fog computing and edge technologies, middleware platforms and the cloud. Information rather than data is increasingly envisioned as the new language of the IoT, while infrastructure and enabling technologies have shifted towards dealing with Big Data use-cases with high scalability or within distributed systems.

Given the traction of big data analytics in the industry and the IoT's potential to become a ``dominant source'' of big data \cite{Chen2014b}, while also a consumer of insights and optimisation drawn from analytics, we foresee that researchers will be looking to understand the process of deriving analytical insights from the IoT. This is further justified by the argument of Akoka \emph{et~al.} \cite{Akoka2017} that ``data of IoT is useful only when analyzed''. As we have noted in our chronological study of previous reviews, this particular combination of areas, with a focus on IoT analytics, to the best of our knowledge, has not been explored in depth. The contribution of this paper is then to:
\begin{enumerate}
\item review IoT analytics applications and research from a variety of domains,
\item propose a classification and taxonomy for IoT analytics to guide future work and
\item review the enabling infrastructure for analytics in the context of big data and examine the tradeoffs to shape research directions.
\end{enumerate}

The methodology used and organisation of the rest of the article is explained next (Section \ref{sec:method}).

\section{Methodology and Organisation of Article}
\label{sec:method}

Section \ref{sec:iot} starts by introducing the IoT vision and application domains, highlighting how this motivates this paper, which is then followed by the main survey content of the paper. The approach employed for the survey follows that of an evidence-based systematic review \cite{Khan2003}. Firstly, two research questions (RQ) were framed:
\begin{enumerate}[label=RQ\arabic*]
\item What IoT analytics research/applications are being published?
\item What enabling infrastructure is required for big data IoT analytics applications?
\end{enumerate} 

Next, we employed an approach of identifying relevant articles through search on the Web of Science platform that indexed an extensive list of multi-disciplinary journals and conferences across multiple databases. The search criteria included the keywords \texttt{`big data'} or \texttt{`analytics'}, filtered by \texttt{`internet of things'}. 460 articles were retrieved from 2011 to 2015. This was updated with 311 articles from 2016 and 2017 when the paper was revised. 

The articles were further screened manually following an inclusion criteria that mandated they \begin{inparaenum}[1)]
\item were from original research,
\item described actual designs, implementations and results,
\item applied analytics and
\item served IoT use-cases.
\end{inparaenum} The highest ranked 6 papers were chosen from each of 5 IoT application domains determined from IoT literature, forming a high-quality pool of 30 papers according to the systematic review method. This addressed RQ1 and is presented in Section \ref{sec:apps}. The ranking was decided proportionately by the number of citations and a qualitative score from 0 to 5 of the technological complexity and completeness of the application (mitigating recency bias).

This understanding of IoT applications was combined with business analytics literature, which has successfully drawn insights from data to optimise business processes, to propose a classification for analytics in Section \ref{sec:analytics}. This classification will help us to better define and target research through an IoT analytics taxonomy as part of the summarisation step of a systematic review.

Finally, we go on to review the current state-of-the-art in IoT infrastructure in Section \ref{sec:infra} that answers RQ2. We used the survey and applications publications previously retrieved on the IoT and identified, by manual inspection, groups of work in cloud, middleware, distributed and fog computing and expanded the search through these keywords to retrieve relevant articles for IoT scenarios as part of the `interpreting the findings' step. Our goal was to consider analytics infrastructure from the perspective of data generation, collection, integration, storage and compute.

The rest of the paper consists of research challenges (Section \ref{sec:challenges}) and a conclusion (Section \ref{sec:conclusion}).

\section{The Internet of Things Vision and Application Domains}
\label{sec:iot}

\subsection{IoT Definition and Common Vision}
\label{subsec:iot_vision}

Both the European Commission and the UK Government Office of Science have a similar vision of the IoT as \quotes{a world in which everyday objects are connected to a network so that data can be shared}, greatly impacting society \cite{UKGovernmentOfficeforScience2014, EuropeanCommission}. The International Telecommunication Union (ITU) calls the IoT \quotes{a global infrastructure for the information society, enabling advanced services by interconnecting things based on existing and evolving interoperable information and communication technologies} \cite{InternationalTelecommunicationUnion2012} and from a broader perspective, \quotes{a vision with technological and societal implications}, which draws its language from a report by the World Economic Forum \cite{WorldEconomicForum2012}.

Common to each of these visions are four principles that are well-defined in IoT literature: 
\begin{enumerate}
\item The IoT exists at a global scale \cite{OvidiuVermesan2013, Lee2013, Vermesan2014},
\item consists of uniquely identifiable Things with sensing or actuating capabilities linked to the physical world \cite{Atzori2010, Kortuem2010, Yick2008},
\item which are interconnected by existing or future technologies so that data can be shared \cite{Miorandi2012, Al-Fuqaha2015}
\item and have potential for societal impact through advanced services \cite{Stankovic2014, Sethi2017, Xu2014}.
\end{enumerate}

The motivation of this paper builds on the third and fourth principles to identify and understand how analytics can enable advanced services from shared and integrated IoT data. The goal then, from these findings, would be to develop various means to help determine what analytics need to be applied and what enabling infrastructure is necessary. First though, we need to define `advanced services'. The next section builds on previous literature to define a set of advanced services domains that help organise the survey of analytics and ensure the paper fulfils a broad coverage.

\subsection{IoT Advanced Services and Application Domains}
\label{subsec:services}

As the IoT develops, many more potential applications and use cases for the IoT will emerge, providing advanced services which offer positive externalities \cite{Holler2014}. A range of advanced service application areas were elicited from each of the surveys describing applications from the 22 in Table \ref{table:surveys} in Section \ref{sec:intro}. They were then classified under their impact to the themes of environment, society and economy which are the drivers of sustainable development used for analysing medium to long term development issues at a large scale \cite{Giddings2002}. Fig. \ref{fig:app_domains} shows the categorisation of the various application areas according to their economic, environmental and societal impact.

\begin{figure}[th]
\centering
\includegraphics[width=0.7\textwidth]{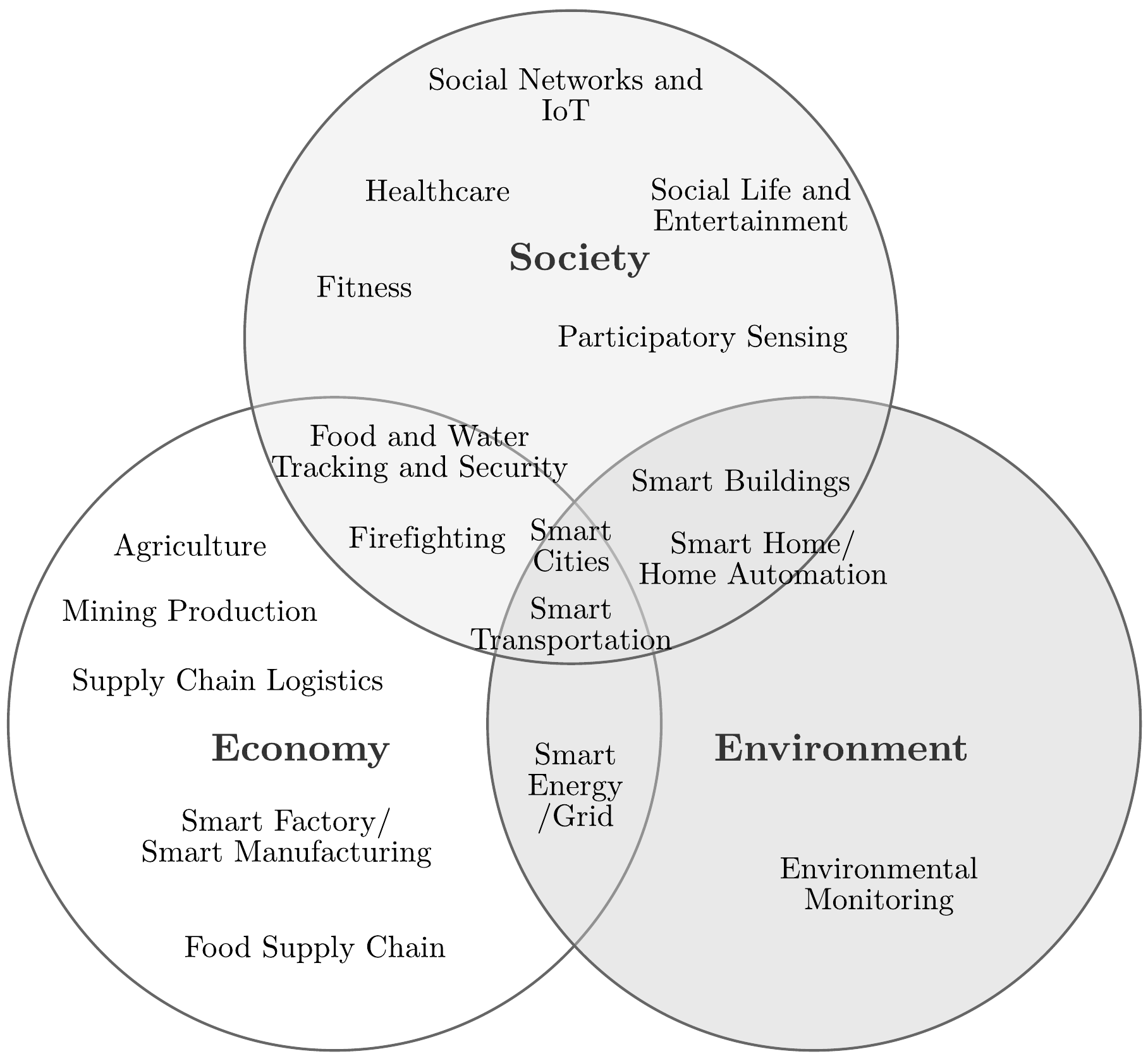}
\caption{Application Areas From Surveys Categorised By Impact to Society, Environment and Economy}
\label{fig:app_domains}
\end{figure}

From these applications areas, a range of application domains including health, transport, living, environment and industry are used to group them, forming the hierarchical classification shown in Fig. \ref{fig:app_domains}. Certain IoT research topics like Smart Cities \cite{Caragliu2011}, Smart Transportation \cite{Sill2011}, Smart Buildings and Smart Homes \cite{Chan2008} which impact multiple themes are also listed.

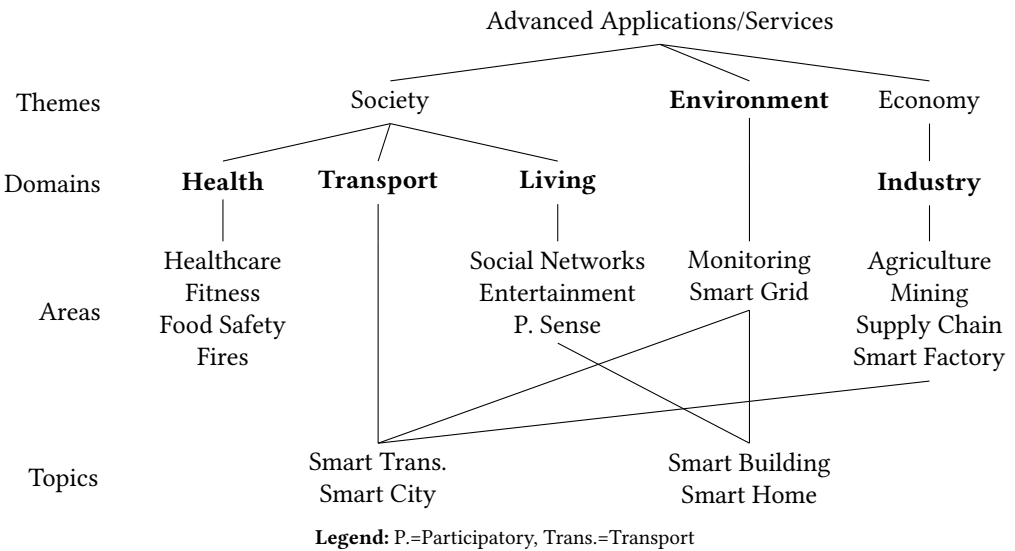
\begin{figure}
\tikzset{every tree node/.style={align=center,anchor=north}}
\begin{tikzpicture}
\centering
\Tree 
[.\node (level0-right) {Advanced Applications/Services}; 
  [.Society 
    [.\node (level2-left){\textbf{Health}};
    	[.\node (level3-left) {Healthcare\\Fitness\\Food Safety\\Fires};
    	]
    ] 
    [.\textbf{Transport} [
    		[.\node [below = 1.6cm] (level4-left) {Smart Trans.\\Smart City};
    		]
    	]
    ] 
    [.\textbf{Living} 
    	[.\node (level3-living) {Social Networks\\Entertainment\\P. Sense};
    	]
    ] 
  ]
  [.\textbf{\textbf{Environment}}  [ 
  		[.\node(level3-centre) {Monitoring\\Smart Grid};
  			[.\node [below = 1.6cm] (level4-right) {Smart Building\\Smart Home};
    		]
  		]
    ] 
  ]
  [.\node (level1-right) {Economy}; 
    [.\node (level2-right) {\textbf{Industry}}; 
    	[.\node (level3-right) {Agriculture\\Mining\\Supply Chain\\Smart Factory};
    	]
    ] 
  ] 
]
\foreach \Value/\Text in {1/{Themes},2/{Domains},3/{Areas}}
{  
  \node[anchor=east] 
    at ([xshift=-1.5cm]{level3-left}|-{level\Value-right}) 
    {\Text};
}
\node[anchor=east] 
    at ([xshift=-3.6cm]{level4-left}) 
    {Topics};
\draw (level3-living.south) -- (level4-right.north);
\draw (level3-right.south) -- (level4-left.north);
\draw (level3-centre.south) -- (level4-left.north);
\end{tikzpicture}
\centering
\footnotesize{\textbf{Legend:} P.=Participatory, Trans.=Transport}
\caption{Application Themes, \textbf{Domains} and Areas Hierachy}
\label{fig:app_domains}
\end{figure}

This classification of advanced service application domains elicited from previous literature serves to advice, organise and ensure the broad coverage of the following survey on IoT analytics applications and infrastructure in Sections \ref{sec:apps}, \ref{sec:analytics} and \ref{sec:infra}.

\section{IoT Applications with Analytics}
\label{sec:apps}

\begin{table}%
\caption{Summary of Analytics Applications by Domains}\label{table:summaryanalyticsapps}
\setlength{\tabcolsep}{-2pt}
\begin{minipage}{\columnwidth}
\begin{center}
\begin{tabular}{cccc}
\toprule
Application & Data Sources & Technique & Currency\\
\midrule
\multicolumn{4}{c}{\textbf{Health}}\\
\midrule
Neo-natal Care \& AAL \cite{Mukherjee2012} 		& $Sn$+$M_{r}$ 	& Data mining, Rules & R\\
AAL \& Navigation \cite{Liu2013} 				& Video 			& Video Analytics & R\\
Smart Clothing Monitoring \cite{Chen2016} 		& $Sn$ 			& Visualisation + ML & H\\
ECG Health Monitoring \cite{Hossain2015}			& $Sn$ 			& Watermark + Classifier & R\\
Prognosis \cite{Ulahannan2002} 					& $Sn$+$M_{r}$ 	& Data mining & H\\
Wellness Recommendations \cite{Banos2016} 		& $Sn$ 			& Classifier + Rules & H\\
\midrule
\multicolumn{4}{c}{\textbf{Transport}}\\
\midrule
Traffic Control \cite{Mak2006}					& Video & Video Analytics & R\\
Pedestrian \& Car Detection \cite{Danner2016}	& $Sn$+Video & Video Analytics + CV & R\\
Behaviour \& Traffic Prediction \cite{Jara2015} & $Sn$ & Visualisation + Model & H\\
Travel Routing \cite{Liebig2014}				& $Sn$ & CRF, A*Search & R\\
Smart Parking \cite{He2014} 					& $Sn$ & Model & R\\
Parking Anomaly Detection \cite{Piovesan2016}	& $Sn$ & Self-Organising Maps & H\\
\midrule
\multicolumn{4}{c}{\textbf{Living}}\\
\midrule
Cultural Behaviour \cite{Chianese2017}			& $Sn$+$S_{m}$ & Visualisation + Model & H\\
Police Situational Awareness \cite{Razip2014}	& $Sn$ & Visualisation & H\\
Public Safety Monitoring \cite{Gimenez2012}		& Video & Video Analytics & R\\
Smart Building Heating \cite{Ploennigs2014} 		& $Sn$ & Anomaly Detection & R\\
Memory Augmentation \cite{Guo2011}				& IoT Data & Data mining & H\\
Wearable Lifestyle Monitor \cite{Mukherjee2014a} & $Sn$ & Anomaly Detection & R\\
\midrule
\multicolumn{4}{c}{\textbf{Environment}}\\
\midrule
Disaster Detection \& Warning \cite{Liebig2014a}& $Sn$+$S_{m}$ & Anomaly Detection & R\\
Urban Disaster Storytelling \cite{Xu2017}		& $S_{m}$ & Data mining & R\\
Wind Forecasting \cite{Mukherjee2013}			& $Sn$ & ANN & H\\
Energy Usage Recommendations \cite{Alonso2013}	& $Sn$ & ML + Rules & H\\
Energy Policy Planning \cite{Ahmed2014} 			& $Sn$ & Classifier + Models & H\\
Smart Energy System \cite{Ghosh2013}				& $Sn$ & Data mining & H\\
\midrule
\multicolumn{4}{c}{\textbf{Industry}}\\
\midrule
On Shelf Availability \cite{Proactive2014}		& Video+$Sn$ & Video Analytics & R\\
SCM Environment Control \cite{Nechifor2014}		& $Sn$+Traffic & CEP & R\\
SCM 4PL \cite{Robak2013} 							& $Sn$ & Ontologies & R \\
Floricultural SCM \cite{Verdouw2013} 			& $Sn$+Traffic & CEP & R\\
Smart Farming \cite{Kamilaris2017}				& $Sn$ & CEP + Ontologies & R \\
Chemical Process Monitoring  \cite{Chiang2017a} 	& $Sn$ & ML (ANN, Gaussian) & R \\
\bottomrule
\end{tabular}
\end{center}
\centering
\footnotesize{\textbf{Legend:} AAL=Ambient Assisted Living, ANN=Artificial Neural Network, CEP=Complex Event Processing, CRF=Conditional Random Fields, CV=Computer Vision, H=Historical, MDP=Markov Decision Process, ML=Machine Learning, $M_{r}$=Medical Records, R=Real-time, SCM=Supply Chain Management, $Sn$=Sensors, $S_{m}$=Social Media}
\end{minipage}
\end{table}%

An important question to ask following our definition of the IoT and its vision is the advantage that connected `things' offer over isolated devices. For example, what is the benefit of deploying a smart parking system as compared to having isolated sensors in a car park using visual signals of green or red on the ceiling to indicate whether a parking lot is empty or occupied? Analytics adds value to integrated data and context from the IoT, producing higher value insights. The analytics-powered smart parking system has a much wider observation space and also guides the user to the available parking lot efficiently, without human intervention, reducing traffic and pollution \cite{Salpietro2015, Bagula2015}. 

Research publications of IoT applications that make use of analytics from 2011 to 2017 were surveyed and the top 6 based on the systematic review methodology (Section \ref{sec:method}) from each application domain introduced in Section \ref{subsec:services} is presented. This is described as follows and summarised in Table \ref{table:summaryanalyticsapps}, which includes the analytics techniques employed, data sources used, and  the currency of the data. Currency refers to whether analytics was applied mainly on historical or real-time data.

\subsection{Health: Ambient Assisted Living, Neo-natal care, Prognosis, Monitoring}

Mukherjee \emph{et~al.} \cite{Mukherjee2012} review the use of data analytics in healthcare information systems. Two analytics applications are Ambient Assisted Living (AAL) \cite{Dohr2010} and neo-natal care. In AAL, rules are applied to IoT data collected from smart objects in the homes of elderly or chronic disease patients while advanced solutions take into consideration contextual information and apply inferencing using ontologies to give health advisories to users, update care-givers or contact the hospital in emergencies. By analysing contextual knowledge in connection with physiological data and being sensitive and adaptive to parameters that vary less frequently, such systems are able to provide descriptive analytics to care-givers and a form of discovery analytics to detect anomalies to trigger emergency warnings. Neo-natal care involves the care of newborn babies where data mining  is applied to multiple data streams to find relationships and patterns and to diagnose any possible medical conditions in infants who are not able to give the doctors verbal feedback.

Analysing the content of video data to aid the elderly and visually handicapped for AAL and navigation respectively is another IoT healthcare application of analytics \cite{Liu2013}.

In their work, Chen \emph{et~al.} \cite{Chen2016} design a smart clothing monitoring system with visualisations of wearable sensor data through a mobile application for use cases like baby, elderly and fitness monitoring. This data is also stored on a `health cloud' integrated with a machine learning library for diagnostic and predictive analytics of medical conditions and users health trends respectively.

Hossain and Muhammad \cite{Hossain2015} show how electro-cardiogram (ECG) and other healthcare data collected from wearable IoT devices and sensors can be watermarked to ensure integrity and sent to the cloud for analysis through feature extraction and classification with a support vector machine (SVM) in real-time. Abnormal patterns are discovered and healthcare professionals alerted.

Analytics can also be applied in the form of prognosis, the science of predicting the future medical condition of a patient, to help healthcare professionals make more informed decisions \cite{Ulahannan2002}. Health indicators collected from sensors of a patient can be compared with data of similar patients and combined with domain knowledge and medical research to make conjectures.

Banos \emph{et~al.} \cite{Banos2016} developed a digital health and wellness framework that collects data streams of IoT health data forming a `life-log' for each user and includes descriptive analytics visualisations of activities. A human activity recogniser combines signal processing, SVM and Gaussian Mixture Models to distinguish activities and recommends activities using rule-based reasoning.

\subsection{Transport: Traffic Control and Routing, Pedestrian Detection, Smart Parking}

Applying analytics on video content has a variety of applications in different fields. In their review paper, Liu \emph{et~al.} \cite{Liu2013} looked at the latest technologies and applications of video analytics and intelligent video systems. Video analytics has been successfully applied in traffic control systems to detect traffic volume for planning, highlighting incidents and enhancing safety by enforcing traffic rules \cite{Mak2006}. Another set of applications is for intelligent vehicles to assist the driver. Danner \emph{et~al.} \cite{Danner2016} introduce their Precedent-Aware Classification (PAC) technique which combines information from previously traveled routes and minimal classification features from sensors to computer vision analytics for pedestrian and car detection on constrained IoT platforms.

Jara \emph{et~al.} \cite{Jara2015} derive insights about human dynamics by analysing the correlation between traffic, temperature and time using IoT sensor data from the SmartSantander smart city testbed \cite{Sanchez2011}. They apply visual analytics to understand and discover insights on human behaviour and use a poisson model to interpolate and predict traffic density. Liebig \emph{et~al.} \cite{Liebig2014} go further by prescribing good routes in travel planning using analytical techniques (a spatiotemporal random field based on conditional random fields \cite{Pereira2001} for traffic flow prediction and a gaussian process model to fill in missing values in traffic data) to predict the future traffic flow and to estimate traffic flow in areas with limited sensor coverage. These were then used to provide the cost function for the A* search algorithm \cite{Hart1968} that uses the combination of a search heuristic and cost function to prescribe optimal routes (provided the heuristic is admissible and predicted costs are accurate).

He \emph{et~al.} \cite{He2014} develop a smart parking service that combines geographic location information, parking availability, traffic and reservation information. The parking process is modelled as a birth-death stochastic process which allows prediction and optimisation of parking availability. Piovesan \emph{et~al.} \cite{Piovesan2016} describe the application of their unsupervised form of self-organising maps (SOM) clustering to the classification of parking spaces according to spatio-temporal patterns. This type of analytics automatically discovers outliers for sensor maintenance and usage anomalies.

\subsection{Living: Cultural Behaviour, Public Safety, Smart Buildings, Memory Augmentation, Lifestyle Monitoring}

Chianese \emph{et~al.} \cite{Chianese2017} describe a system for cultural behaviour analysis. They combine models and proximity evaluation algorithms to classify movement in museums from sensors with semantic enrichment from knowledge bases of cultural exhibits and social media of cultural tourism to analyse cultural behaviour using visualisations within an associative model.

Visualisation that taps the human cognitive ability to recognise patterns has also been employed by Razip \emph{et~al.} \cite{Razip2014} in helping law enforcement officers increase their situational awareness. Officers are equipped with mobile devices that tap into crime data and spatio-temporal sensor data to show interactive alerts of hotspots, risk profiles and on demand chemical plume models.

Additionally, there are public safety and military applications that apply video analytics in detecting movement, intruders or targets. The public safety use case is elaborated on by Gimenez \emph{et~al.} \cite{Gimenez2012} where they discuss how given the big data problem of having huge amounts of video footage, smart video analytics systems can proactively monitor, automatically recognise and bring to notice situations, flag out suspicious people, trigger alarms and lock down facilities through the recognition of patterns and directional motion, recognising faces and spotting potential problems by tracking, with multiple cameras, how people move in crowded scenes.

Ploennigs \emph{et~al.} \cite{Ploennigs2014} show how analytics can be applied to energy monitoring used in heating for smart buildings. The system is able to diagnose anomalies in the building temperature, for example, break downs of the cooling system, high occupancy of rooms, or open windows causing air exchange with the external surroundings. Using a semantics based approach, the Building Automation and Control Systems (BACS) \cite{Aste2017} could, from the sensor definitions, automatically derive diagnosis rules and behaviour of a specific building, making it sensitive to new anomalies.

Guo \emph{et~al.} \cite{Guo2011} look at discovering various insights from mining the digital traces left by IoT data from cameras, wearables, mobile phones and smart appliances. Resulting applications are life logging systems to augment human memory with recorded data, real world search for objects and interactions with people and a system to improve urban mobility systems by studying large-scale human mobility patterns.

Mukherjee \emph{et~al.} \cite{Mukherjee2014a} present a fast algorithm for detecting anomalies and also for classifying high dimensional data. These were tested with accelerometer data from a wearable personal digital assistant to recognise human activity in real time but can be generalised to other types of high dimensional data. The importance of such algorithms in detecting anomalies and discovering patterns to classify activity from sensor data are analytical tools that form a basis for smart and intelligent devices and in this example, for activity tracking and monitoring.

\subsection{Environment: Disaster Detection \& Response, Wind Forecasting, Smart Energy}

Schnizler \emph{et~al.} \cite{Liebig2014a} describe a disaster detection system that works on heterogenous streams of sensor data. Their method includes Intelligent Sensor Agents (ISAs) that produce anomalies, low level events with location and time information e.g. an abnormal change in mobile phone connections at a ISA in a telecom cell or base station, a sudden decrease in traffic, increase in twitter messages, change in water level or change in the volume of moving objects at a certain location. These anomaly events then enter Round Table (RT) components that fuse heterogenous sources together by mapping them to a common incident ontology through feedback loops that might involve crowdsourcing, human-in-the-loop or adjusting parameters of other ISAs to find matches. The now homogenous incident stream, can then be processed by a Complex Event Processing (CEP) \cite{Luckham2002} engine to complete the situation reconstruction by doing aggregation and clustering with higher-level semantic data, simulation and prediction of outcomes and damage. The resultant incident stream can provide early warning effecting early disaster response. 

Xu \emph{et~al.} \cite{Xu2017} also present a disaster detection system targeted instead at urban disasters. They utilise social media events from multi-modal microblog posts (videos, images and text) to mine semantic, spatiotemporal and visual information producing a story. This real-time story of urban emergencies unfolding serves to increase the situational awareness of emergency response teams.

Another environmental application is wind forecasting \cite{Mukherjee2013}. Data is collected from wind speed sensors in wind turbines and an Artificial Neural Network is used on this data and historical data to perform the forecasting. This is useful for energy provision and planning.

Ghosh \emph{et~al.} \cite{Ghosh2013} have implemented a localised smart energy system that uses smart plugs and data analysis to actively monitor energy policy and by performing pattern recognition analysis on accumulated data, spot additional opportunities to save energy. This resulted in saving on electricity bills especially by reducing the amount of power wasted in non-office hours from appliances, desktops and printers.

Similar work by Alonso \emph{et~al.} \cite{Alonso2013} works on using machine learning and an expert system (rule-based) to provide personalised recommendations, based on energy usage data collected in Smart Homes, that help a user to more efficiently utilise energy. They go one step further to provide recommendations through predicting cheaper options by detecting similar patterns in big data collected from other homes. Ahmed \cite{Ahmed2014} applies similar analysis on combined consumption data for use in organisations to help in energy policy planning. He develops a model to classify the energy efficiency of buildings and the seasonal shifts in this classification and using more detailed appliance specific data, forecasts future energy usage.

\subsection{Industry: Supply Chain Management, Smart Farming, Chemical Process}

Vargheese \emph{et~al.} \cite{Proactive2014} propose a system that improves shoppers' experience by enhancing the On the Shelf Availability (OSA) of products. Furthermore, the system also looks to forecast demand and provide insights on buyers' behaviour. A multi-tiered approach is employed, where sensors like video cameras, process video streams locally and analyse the products on the shelf, this data is then verified by other sensors like light, infra-red and RFID sensors and the metadata produced is sent to the the cloud to be further processed. In the cloud, this real time data is combined with models from learning systems, data from enterprise Point of Sale (POS) systems and inventory systems to recommend action plans to maintain the OSA of products. The staff of the store are informed and action is taken to restock products. Weather data, local events and promotion details are then analysed with the current OSA to provide demand forecasting and to model buyers behaviour which is fed back into the system. 

Nechifor \emph{et~al.} \cite{Nechifor2014} describe the use of real time data in analytics in a cold chain monitoring \cite{Abad2009} process. Trucks are used for transporting perishable goods and drugs that require particular thermal and humidity conditions, sensors measure the position and conditions in the truck and of each package, while actuators - air conditioning and ventilation can be controlled automatically. On a larger scale, predictions can be made on delays in routes and when necessary to satisfy the product condition needs, longer but faster routes (less congestion) might be selected. 

Similarly, Verdouw \emph{et~al.} \cite{Verdouw2013} and Robak \emph{et~al.} \cite{Robak2013} examine supply chains - the integrated, physical flow from raw material to end products with a shared objective, and formulate a framework based on their virtualisation in the IoT. At its highest level, a virtual supply chain supports intelligent analysis and reporting. This is applied to floricultural and a Fourth Party Logistics (4PL) integrator respectively, where business intelligence, data mining and predictive analytics can provide early warning in case of disruptions or unexpected deviations and advanced forecasting about consequences of the detected changes when the product reaches destination.

In the above examples on product and supply chain management, we see a common theme of predictive analytics being employed to business processes. This predictive analytics is often powered by learning from data to discover models or through data mining for patterns in data. The effectiveness of these algorithms benefits from the big data of the IoT in providing a large observation space for discovering patterns and trends. Real time data from sensors then provide the information required to immediately control actuators to rectify problems like products being out of stock on the shelf or conditions in trucks being unsuitable for perishable food. 

Kamilaris \emph{et~al.} \cite{Kamilaris2017} describe the use of a Complex Event Processing (CEP) engine to discover significant events on semantically-enriched data streams from sensors within two smart farming scenarios. One scenario included detecting the fertility of cows from temperature readings and other information on a dairy farm to suggest the best insemination timings. The other was to adaptively control the soil conditions for crop cultivation.

The chemical process industry deploys inferential industrial IoT sensors to process monitoring chains \cite{Chiang2017a}. Some techniques applied by sensors include linear regression, artificial neural networks (ANN) and Gaussian process regression which predict variables using available process data. These predictions enable quality monitoring and advance control systems in plants to automatically react and prescribe process modifications ``to prevent off-grade products''.

\section{Types of Analytics and their importance}
\label{sec:analytics}
Following the study of the current work in analytics in the IoT, we explore a classification of analytics that is applicable to these domains. We derive a categorisation of analytical capabilities from business analytics literature, which the term analytics comes from. Bertolucci \emph{et~al.} \cite{Bertolucci2013} propose descriptive, predictive and prescriptive categories while Gartner \cite{Kart2012} \cite{Chandler2011} proposes the extra category of diagnostic analytics. Finally, Corcoran \emph{et~al.}  \cite{Maitland2012} introduce the additional category of discovery analytics. We build upon these to form a comprehensive classification of analytic capabilities consisting of five categories: descriptive, diagnostic, discovery, predictive and prescriptive analytics. Each category is described in detail in Section \ref{subsec:analytics_cap} and we also summarise how each IoT application surveyed in the previous section is categorised in Table \ref{table:summaryapps}. Each application domain has applications which support multiple analytical capabilities. We also note that all the categories of capabilities are well-represented in the literature survey, while mature domains like the industrial IoT focus on high value analytics.

\begin{table}
\centering
\caption{Summary of Application References by their Domains and Analytical Capabilities}
\label{table:summaryapps}
\begin{tabular}{l|p{1.5cm}|p{1.7cm}|p{2.2cm}|p{1.9cm}|p{1.9cm}}
\toprule
\theadfont\diagbox[width=9em]{Domain}{Capability} & \centering{}Descriptive & \centering{}Diagnostic & \centering{}Discovery & \centering{}\centering{}Predictive & {\centering Prescriptive\\}\\ 
\midrule
Health & \cite{Mukherjee2012,Chen2016} &  \cite{Chen2016, Liu2013} & \cite{Mukherjee2012, Hossain2015, Banos2016} &  \cite{Ulahannan2002} &  \cite{Chen2016}\\
\midrule
Transport & \cite{Mak2006} & & \cite{Danner2016, Piovesan2016, Jara2015} & \cite{He2014}  & \cite{Liebig2014} \\
\midrule
Living &\cite{Razip2014, Chianese2017} & \cite{Ploennigs2014} & \cite{Gimenez2012, Chianese2017, Guo2011, Mukherjee2014a} &  & \\
\midrule
Environment &  \cite{Xu2017} & & \cite{Liebig2014a, Ghosh2013} & \cite{Mukherjee2013,Alonso2013, Ahmed2014} & \\
\midrule
Industry & & & \cite{Verdouw2013, Robak2013} & \cite{Nechifor2014, Verdouw2013, Chiang2017a} & \cite{Proactive2014, Kamilaris2017, Chiang2017a} \\
\bottomrule
\end{tabular}
\end{table}

\begin{figure}[!h]
\centering
\includegraphics[width=3.4in]{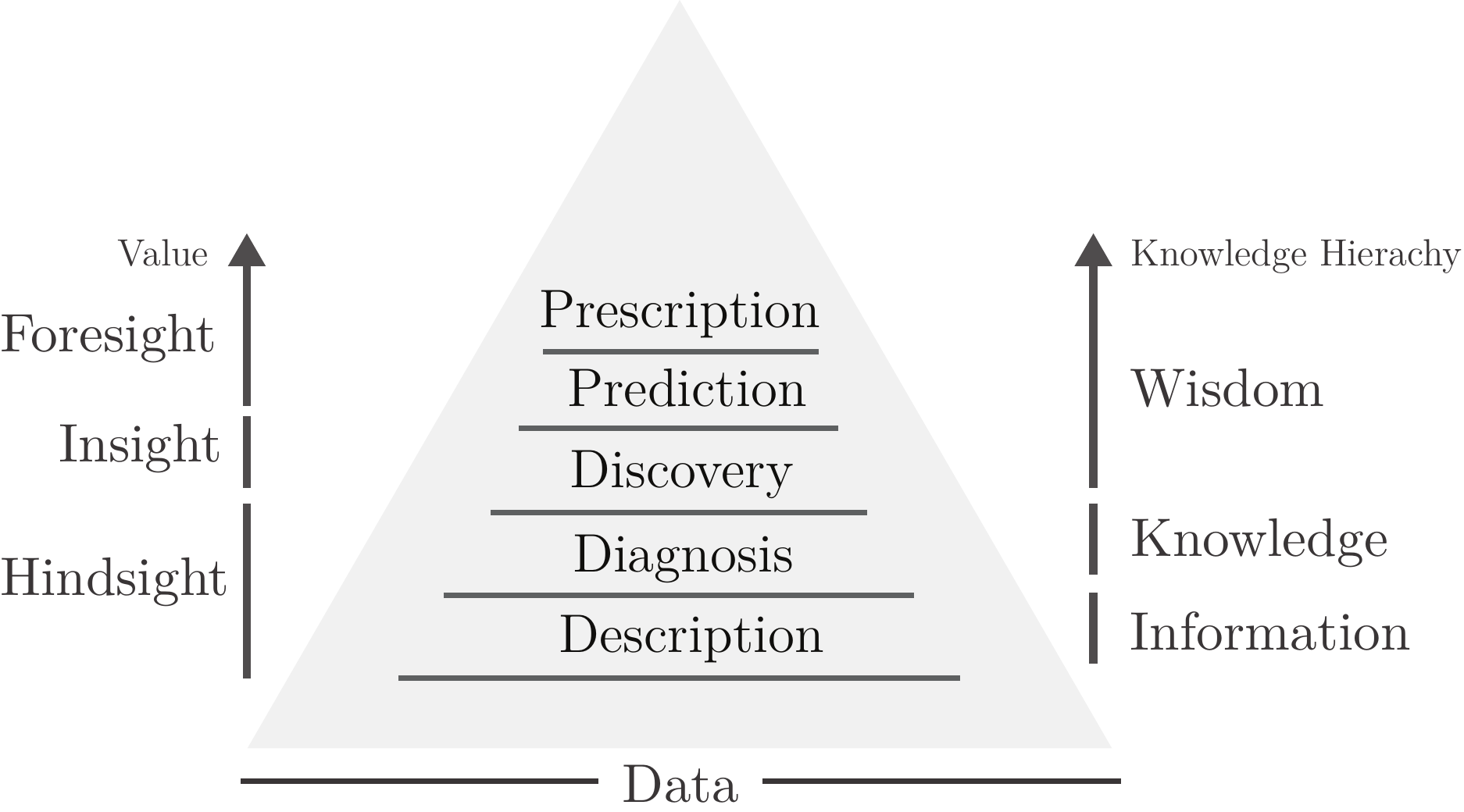}
\caption{Analytics and the Knowledge and Value Hierachies}
\label{fig:knowledgehierachy}
\end{figure}

Fig. \ref{fig:knowledgehierachy} looks at how each analytical capability fits within the Knowledge Hierarchy \cite{Bernstein2011} which is a common framework used in the Knowledge Management domain. This categorisation of analytic capabilities enables us to establish what the aim of analysis is and allows us to relate to the vision of IoT deployment as often expressed in research roadmaps. The value of each capability, is also highlighted in the figure. The knowledge hierarchy starts with data at the base, examples of which are facts, figures and observations (e.g. the raw data produced by IoT 'things'). Information is interpreted data with context, for example, temperature as represented by descriptive analytics: an average over a month or a categorical description of the day being sunny and warm. Knowledge is information within a context with added understanding and meaning, perhaps possible reasons for the high average temperature this month. Finally, wisdom is knowledge with insight, for example, discovering a particular trend in temperature and projecting it across future months while providing cost saving energy management solutions for heating a smart home based on these predictions. Each component of the knowledge hierarchy builds on the previous tier and we can see something similar with analytical capabilities. To add a practical view from business management literature to our discussion, a review of organisations adopting analytics \cite{Lavalle2010} categorised them as Aspirational, Experienced and Transformed. Aspirational organisations were seen to use analytics in hindsight as a justification for actions, utilising the data, information and knowledge tiers in the process. Experienced organisations utilised insights to guide decisions and transformed organisations were characterised by their ability to use analytics to prescribe their actions, effectively applying foresight in their decision making process.

\subsection{Five Categories of Analytics Capabilities}
\label{subsec:analytics_cap}
\subsubsection{Descriptive Analytics}
It helps us to answer the question, ``what happened?''. It can take the form of describing, summarising or presenting raw IoT data that has been gathered. Data are decoded, interpreted in context, fused and then presented so that it can be understood and might take the form of a chart, a report, statistics or some aggregation of information. 
\subsubsection{Diagnostic Analytics}
It is the process of understanding why something has happened. This goes one step deeper then descriptive analytics in that we try to find out the root cause and explanations for the IoT data. Both descriptive and diagnostic analytics give us hindsight on what and why things have happened.
\subsubsection{Discovery in Analytics}
Through the application of inference, reasoning or detecting non trivial information from raw IoT data, we have the capability of Discovery in Analytics. Given the acute problem of volume that big data presents, Discovery in Analytics is also very valuable in narrowing down the search space of analytics applications. Discovery in Analytics on data tries to answer the question of what happened that we don't know about and the outcome is insight into what happened. What differentiates this from the previous types of analytics is using the data to detect something new, novel or different (e.g. trends, exceptions or clusters) rather than describing or explaining it.
\subsubsection{Predictive Analytics}
For the final two categories of analytics, we move from hindsight and insight to foresight. Predictive Analytics tries to answer the question: ``what is likely to happen?''. It uses past data and knowledge to predict future outcomes \cite{Jr2007} and provides methods to assess the quality of these predictions \cite{Alpren2010}.
\subsubsection{Prescriptive Analytics}
It looks at the question of what should I do about what has happened or is likely to happen. It enables decision-makers to not only look into the future about opportunities (and issues) that are potentially out there, but it also presents the best course of action to act on foresight in a timely manner \cite{Basu2013} with the consideration of uncertainty. This form of analytical capability is closely coupled with optimisation, answering `what if' questions so as to evaluate and present the best solution.

\subsection{Specific Types of Analytics}
\begin{figure}[!t]
\centering
\includegraphics[width=0.7\textwidth]{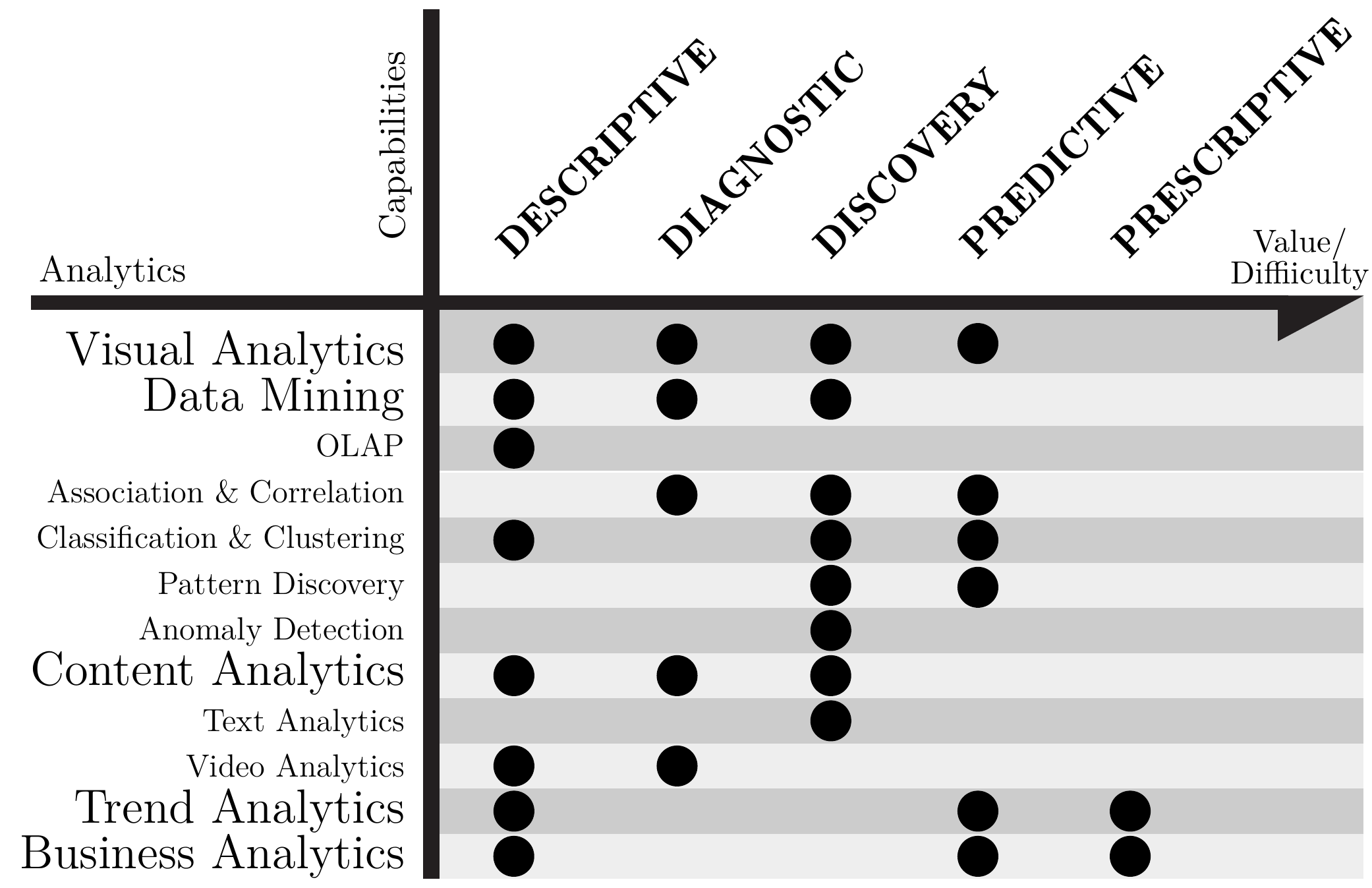}
\caption{Classification of Types of Analytics}
\label{fig:analytics}
\end{figure}

Having looked at analytical capabilities which help to define the aims of analytics, we look at specific analytics that can guide stakeholders involved in the deployment of analytics on IoT applications. A summary of the specific types of analytics and their corresponding analytical capabilities can be found in Fig. \ref{fig:analytics}.

\subsubsection{Visual Analytics}
Visual analytics combines interactive visualisations with data analytics techniques  \quotes{for an effective understanding, reasoning and decision making on the basis of very large and complex data sets} \cite{Keim2008}. Hence, visual analytics can contribute to not only describing and diagnosing what happened but also help users to discover new insights. In the work by Zhang et. al \cite{Zhang}, we see visual analytics being applied to health care data and describing, through answering of questions like ``What is the distribution of pregnancy age?'', diagnosing, through hypothesising two disease patterns due to ``diarrhoea'' and ``fever'' not being correlated and discovery, through detecting the delayed outbreak of two diseases.
\subsubsection{Data Mining}
Data Mining is part of the Knowledge Discovery from Data (KDD) process in which interesting patterns and knowledge are discovered from large amounts of data \cite{Han2012}. The IoT is a source for a large amount of data in which the techniques of data mining can be applied. These include:

\textit{Multi-dimensional data summary} is often associated with Online analytical processing  (OLAP) operations that make use of background knowledge of the domain to allow presentation of data at different levels of abstraction. For example, you could drill-down and roll-up data to present it at different degrees of summarisation.

\textit{Association \& correlation} is the process of finding the relationship between two variables which vary according to some pattern. This could allow us to find out whether buying product A, led to buying product B with a degree of confidence and support.

\textit{Classification} is the process of finding some model or function that has the ability to distinguish between data classes or concepts.

\textit{Clustering} is the process of grouping data objects into classes without labels. The clustered data objects have maximum similarity to in-class objects and minimum similarity between objects from other classes.

\textit{Pattern discovery} is the process of detecting and extracting interesting patterns from data, an example of which are frequent item sets, a set of items that often appear together in a transactional data set. \textit{Anomaly detection} refers to the problem of \quotes{finding patterns in data that do not conform to expected behaviour} \cite{Chandola2009}.

\subsubsection{Content and Text Analytics}
Content Analytics is the broad area of which analytical techniques are applied to digital content. Text analytics is the derivation of high quality information from unstructured text, for example, extracting named entities and relations, analyse sentiment, extract events and time series information, etc.
\subsubsection{Video Analytics}
Video Analytics (VA) is about the use of specialised software and hardware \quotes{to analyse captured video and automatically identify specific objects, events, behaviour or attitudes in video footage in real-time} \cite{Gimenez2012}.
\subsubsection{Trend Analytics}
Trend analytics is concerned with looking at data and events across time, understanding it and making predictions to future trends and providing early warning systems. Trend analytics is also closely related to the analysis of time-series information \cite{Chatfield2013}, where looking at a time-series we try to find a `long-term change in the mean level'.  
\subsubsection{Business Analytics}
Business Analytics is the practice of using an organisations data to gain insights through analytical techniques that can better inform business decisions and automate and optimise business processes. \\

\subsection{A Layered Taxonomy of Data, Analytics and Applications for the IoT}

\begin{figure}[ht]
\centering
\includegraphics[width=\textwidth]{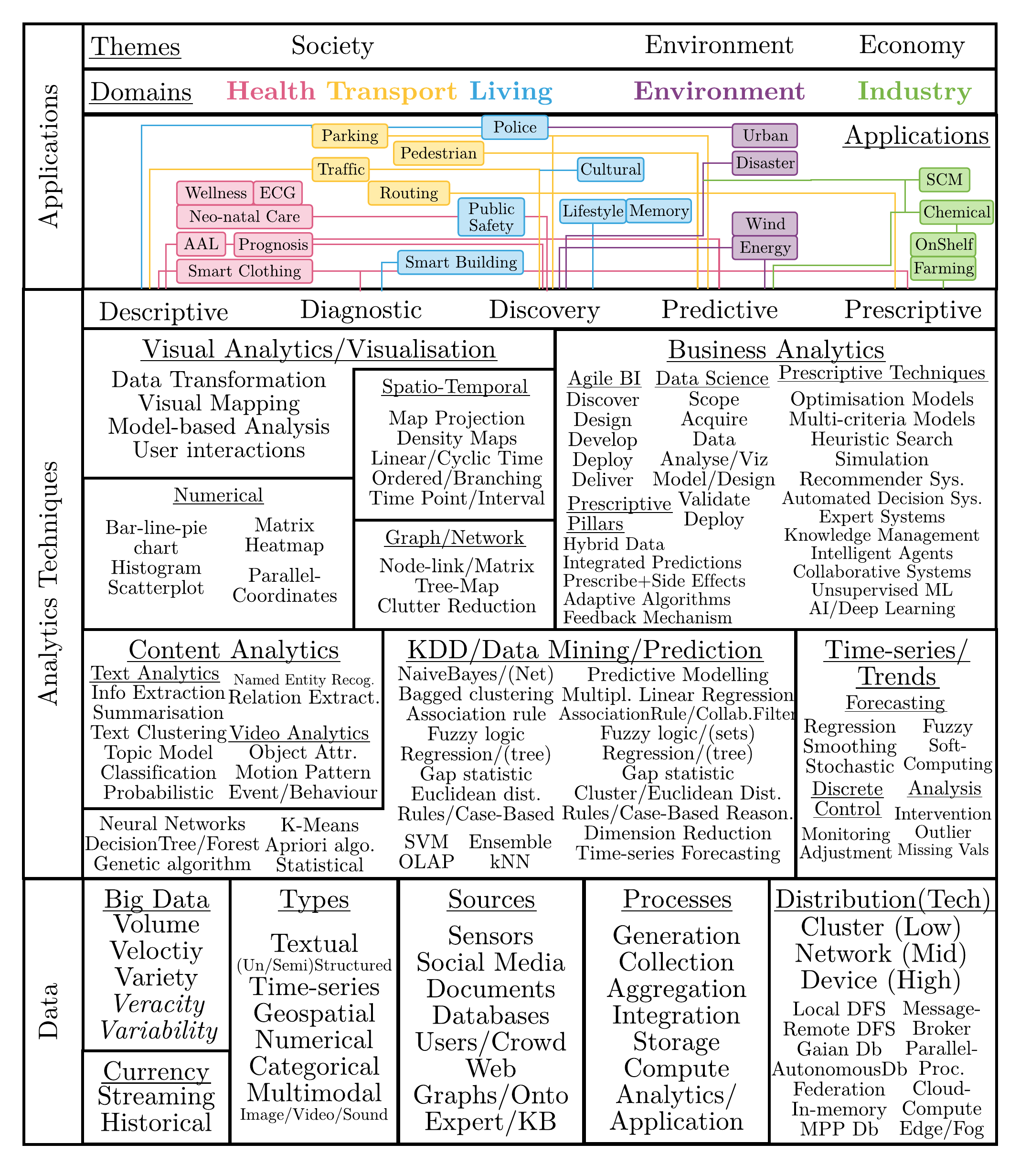}
\caption{Layered Taxonomy of Analytics From Data to Application}
\label{fig:analytics_layers}
\end{figure}

Fig. \ref{fig:analytics_layers} shows a layered taxonomy of analytics for the IoT that summarises our survey with respect to analytics capabilities and specific analytics. There are three layers in the taxonomy: data, analytics and applications. Within each layer are various concepts, classes and techniques which are well-defined in background literature and gathered from reviews in each area.

In the analytics layer, visual analytics processes are defined by Keim \emph{et~al.} \cite{Keim2010} while techniques for each data type are summarised in surveys \cite{Zhang2012,Sun2013,Aigner2008}. Data mining \cite{Liao2012,Goebel1999,shmueli2017data}, text analytics \cite{aggarwal2012mining} and video analytics \cite{Liu2013} each are well-described in the referenced authoritative texts. Time-series forecasting \cite{Mahalakshmi2016}, analysis and control \cite{box2015time} have also been reviewed in detail. Literature also covers business analytics processes \cite{Larson2016}, prescriptive analytics \cite{Basu2013} and techniques \cite{sharda2014businesss}.

In the application layer, themes and domains are from Section \ref{subsec:services} while the IoT applications from each domain surveyed in Section \ref{sec:apps} are shown connected to their various analytics capabilities. Analytics techniques can then be referenced under each capability. 

In the data layer, big data as defined in Section \ref{sec:intro} is summarised along with terms used throughout the survey including currency, types of data and their sources. Two other terms for big data, veracity and variability are introduced for completeness. Veracity is concerned with the noise within data and how accurate the data is for whatever purpose it is to serve. Variability is concerned with data whose meaning changes due to differences in interpretation of data within a specific context. Finally, processes, distribution levels and distributed technologies for storage and compute are covered in Section \ref{sec:infra} that follows this.

\section{Enabling Infrastructure for IoT Analytics}
\label{sec:infra}

In the previous section we looked at classifying analytics and building a taxonomy for understanding analytics. In this section, we will review work that enables analytics to be applied on IoT data. 

\begin{figure}
\centering
\includegraphics[width=0.7\textwidth]{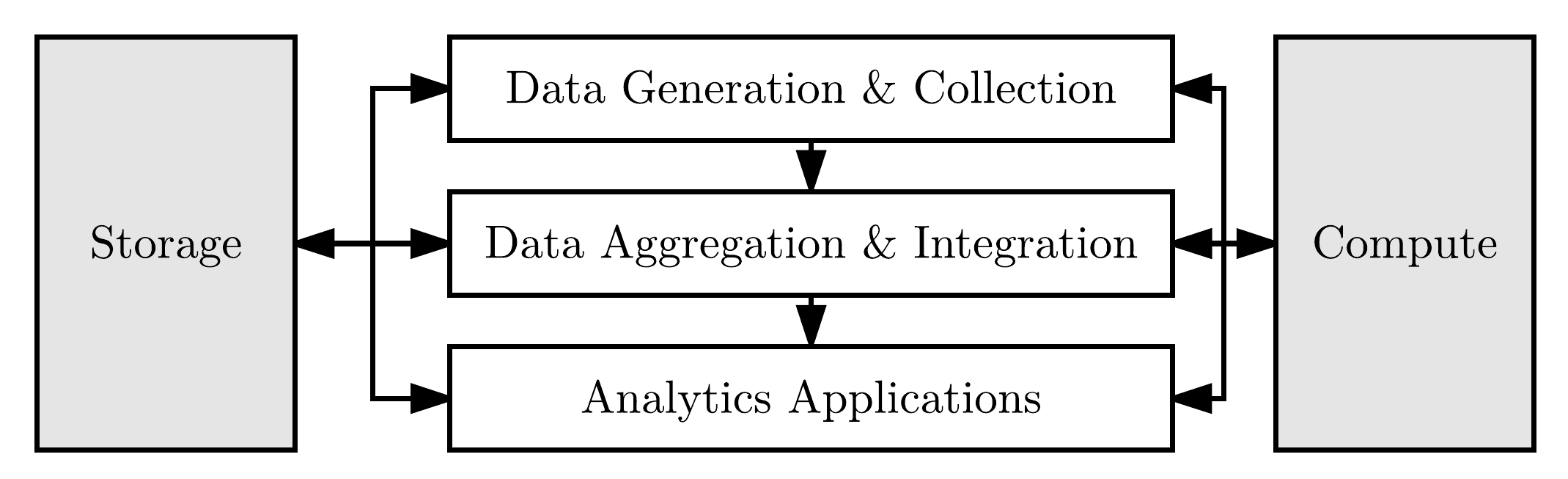}
\caption{Data Flow Process for Analytics Applications}
\label{fig:dataflow}
\end{figure} 

Enabling infrastructure for analytics on the IoT are components, techniques and technology that contribute to the process whereby data is utilised in analytics applications. Fig. \ref{fig:dataflow} shows the process of how data goes through the steps of generation and collection, aggregation and integration and finally is applied in analytics applications \cite{Chen2014a}. Storage and compute are abstract processes involved with each step of this data flow. In practice, data could be pipelined from one step to another, hence, need not necessarily be stored, physically, in a separate location. Compute could also be done on the device or in transit and need not imply a separate compute component. 

\begin{figure}
\centering
\includegraphics[width=0.9\textwidth]{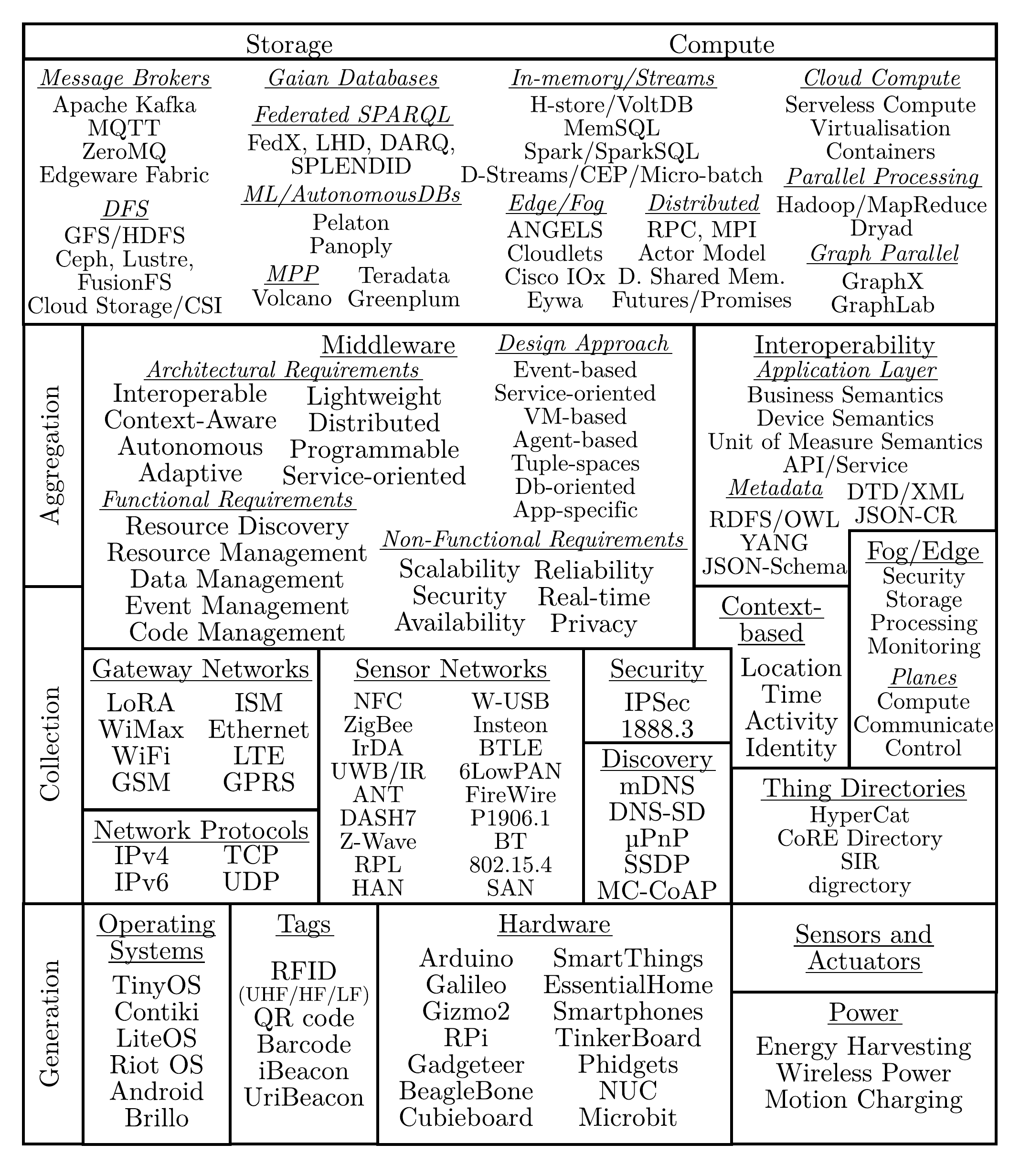}
\caption{IoT Enabling Infrastructure for Analytics: Generation, Collection, Aggregation, Storage, Compute}
\label{fig:enabling_infra}
\end{figure} 

The following sections elaborate on each step of the data flow in IoT analytics from data generation and collection to aggregation and integration with storage and compute alongside. Fig. \ref{fig:enabling_infra} summarises the technologies covered.

\subsection{Data Generation: Sensors and Tags, Hardware and OS, Power}

A major source of data in the IoT is generated from sensors including many types of environmental, spatial sensors and health sensors \cite{Ray2015}. Tags also generate data and can be passive like QR code and barcode patterns which require a device to scan or be active like iBeacon \cite{Apple2017} and UriBeacon \cite{Google2017} technologies which project signals to mobile applications. RFID tags can be either passive or active, the active type requiring a power source to broadcast signals, and can be UHF (Ultra High Frequency), HF (High Frequency), or LF (Low Frequency). A list of hardware platforms for sensors or base stations receiving the generated data and a list of lightweight operating systems for the IoT are discussed in the surveys by Ray \cite{Ray2016} and Razzaque \emph{et~al.} \cite{Razzaque2016} respectively.

Remotely-deployed IoT sensors also require power especially for the energy consuming process of wirelessly transmitting data. Wolf \cite{Wolf2017} describes a number of energy scavenging systems that harvest energy from the environment, while wireless charging technologies like ubeam \cite{Ubeam2017} and motion charging like Ampy \cite{Ampy2017} are alternatives. Data is then transmitted and collected as follows.

\subsection{Data Collection: Discovery, Management, Transmission, Context and Fog}

A significant amount of work on the IoT has been to develop middleware, the software layer that connects various components like the device, storage, compute and network together. Middleware in the IoT has functional requirements \cite{Razzaque2016,SomaBandyopadhyay2011} including: 
\begin{inparaenum}[1)]
\item resource discovery, 
\item resource management, 
\item data management, 
\item event management and
\item code management.
\end{inparaenum} Of these requirements, resource discovery and management fit within the collection step while data and event management fit within the aggregation and storage processes while code management fits within the compute process. 

There are a number of technologies for the IoT that support the discovery of devices, Multicast DNS (mDNS) \cite{ietf2013}, DNS Service Discovery (DNS-SD) \cite{Cheshire2017}, Micro Plug and Play ($\mu$PnP) \cite{Yang2015}, Simple Service Discovery Protocol (SSDP) \cite{ietf1999} and Multicast CoAP (MC-CoAP) \cite{ietf2014}. One means of managing the discovered resources is through Thing Directories that serve as catalogues of resources. HyperCat \cite{Alliance2017}, CoRE Resource Directory \cite{ietf2017}, Sensor Instance Registry (SIR) \cite{jirka2010} and digrectory \cite{Jara2013} are various implementations supporting resource lookup and search.

Another important process in data collection is the transmission of generated data. We divide the transmission technologies into those for communication within sensor networks like Zigbee and those for communication within gateway networks and the wider IoT like LTE and GSM. These technologies are discussed in the survey by Ray \cite{Ray2015} on IoT architecures. Network and transport layer protocols like IPv4/v6 and TCP/UDP are well-defined in literature. IPSec \cite{InternetEngineeringTaskForce2005} is a security protocol suite for the network layer that authenticates and encrypts packet data while 1888.3 \cite{IEEE2013} is a security standard for the IEEE Ubiquitous Green Community Control Network.

Context-based computing is a research area within the IoT that involves the detection, sharing and grouping of devices according to context in the IoT. Context from the conceptual perspective, as described by Perera \emph{et~al.} \cite{Perera2014}, refers to the location, time, activity and identity related to data collected. Grim \emph{et~al.} \cite{Grim2011} design a bloom filter \cite{Bloom1970} inspired data structure that summarises this context and identifies set membership in a probabilistic way so resources can be discovered and grouped. Perera \emph{et~al.} \cite{Perera2013} implement resource search and management on a context-based framework. A Comparative Priority-based Weighted Index is generated for each resource, combining priorities like accuracy, reliability, energy, cost and availability which optimises the selection process for the aggregation of data sources.

Chiang \emph{et~al.} \cite{Chiang2017} define fog computing as an ``end-to-end horizontal architecture'' for the IoT that distributes the compute and storage, control and communication planes nearer to users ``along the cloud-to-thing continuum''. Aazam and Huh \cite{Aazam2014} describe specifically how this vision can be realised in terms of additional security, storage, processing and monitoring sub-layers between the physical layer and the transport layer of an IoT architecture. Hence, Fog Computing extends to the data aggregation layer and can even extend to the analytics process.

\subsection{Data Aggregation and Integration: Interoperability}

Besides the functional requirements of middleware defined in the previous section, the survey by Razzaque \emph{et~al.} \cite{Razzaque2016} also describes architectural requirements, design approaches and non-functional requirements of middleware, as shown in Fig. \ref{fig:enabling_infra}, along with a detailed review of various software and publications. Interoperability is one of the architectural requirements and is essential for the data aggregation and integration process. McKinsey \cite{Manyika2015} estimate that such interoperability will unlock an additional 40 to 60 percent of the total projected future IoT market value.

Berrios \emph{et~al.} \cite{Berrios} describe how various cross-industry consortia concerned with the IoT are converging on semantic interoperability within the application layer which they split into interoperability of business semantics, device semantics, unit of measure semantics and API and service standards. All the consortia involved are working on device semantics for interoperability while at least one consortium has defined standards for each of the home \& buildings, retail, healthcare, transport \& logistics and energy industries. The series of articles, co-authored by representatives from each consortia, also recommended a top-level ontology, an ontology representing the intersection of business and device semantics and a common data format.

Milenkovic \cite{Milenkovic2015} also argue for a common representation for metadata, that provides context to the data collected. Linked Data, which is defined as \quotes{a set of best practices for publishing data on the Web so that distributed structured data can be interconnected and made more useful by semantic queries} \cite{Bizer2009}, is seen as one means. Barnaghi \emph{et~al.} \cite{Barnaghi2012} argue that Linked Data and semantic technologies can serve to facilitate interoperability, data abstraction, access and integration with other cyber, social or physical world data. RDFS, which inspired the popular schema.org vocabulary that allows persons, events, places and products to be defined on the web and the Web Ontology Language (OWL) for complex modelling and non-trivial automated reasoning in ontologies are related technologies that allow metadata to be represented. There are also proposals for other data models like YANG \cite{schonwalder2010network}, JSON Schema \cite{galiegue2013json} and JSON Content Rules \cite{cordell2016language} to be adopted.

\subsection{Architectures for Storage and Compute}
\label{subsec:archi}

At a high level, architectures help to define how to build infrastructures and how to handle big IoT data in the storage and compute components for analytics. One such architecture is the lambda architecture by Marz \emph{et~al.} \cite{Marz2014} which consists of a speed, a serving and a batch layer. The idea is that for huge datasets it is necessary to precompute batch views in the batch layer and update them in the serving layer, at the same time a speed layer compensates for the high latency of the batch computations by looking at recent data and doing fast incremental updates. 

This big data architecture is useful in providing us with a general idea of how analytics can scale to the volume of IoT data. Ye \emph{et~al.} \cite{Ye2013} implement a service for big data analytics (using R and Hadoop for efficient parallel processing \cite{Das2010}) in the batch layer to do data mining tasks like clustering. Products like Onix \cite{Shtykh2014}, which do analytics on streams, work on implementing solutions for the speed layer while industry players like MapR \cite{MapRTechnologies2014} have also proposed the Lambda Architecture as part of their data processing architecture. The Lambda Architecture has also been used in an IoT context by Villari \emph{et~al.} \cite{Villari2014}, who apply it to a Smart Environment use case. 

Baldominos \emph{et~al.} \cite{Baldominos2014} also propose a design that is similar in structure to the Lambda architecture and is another example of how an analytics system, for doing machine learning and recommendations in this case, can be implemented with this separation of batch (batch machine learning module/storage), speed (stream machine learning module) and serving (dashboard) layers.

The Hadoop and Spark ecosystems are two other big data processing architectures. Hadoop consists of two main parts, a Distributed File System (DFS) like HDFS and a distributed programming model like MapReduce. The Hadoop ecosystem\footnote{Available from \href{http://thebigdatablog.weebly.com/blog/the-hadoop-ecosystem-overview}{http://thebigdatablog.weebly.com/blog/the-hadoop-ecosystem-overview}} consists of various technologies built on top and around these two parts including warehousing like Hive, NoSQL databases like HBase, data ingestion pipes like Flume and machine learning libraries like Mahout and a host of other technologies\footnote{Available from \href{https://hadoopecosystemtable.github.io/}{https://hadoopecosystemtable.github.io/}}. 

The Spark ecosystem is built on Spark Core and consists of components like SparkSQL, Spark Streaming, MLLIB and GraphX amongst others. Spark is described in more detail in Section \ref{subsec:compute_tech}.

The Lambda architecture, Hadoop and Spark ecosystems, however, are suited for big data systems in which compute and storage are in centralised or cloud-based clusters rather than decentralised fog and edge based computing. The next two sections describe storage and compute technologies which can be used for the IoT and big data analytics including fog computing technologies. Table \ref{table:products} summarises the distributed storage and compute technologies and their references.

\begin{table}
\setlength{\tabcolsep}{4pt}
\caption{Summary of Distributed Storage and Compute Technologies}\label{table:products}
\begin{minipage}{\columnwidth}
\begin{center}
\begin{tabular}{lll}
\toprule
Technology & Product & Remark\\
\midrule
Locally managed DFS & GFS/HDFS \cite{Ghemawat2003} 	& Centralised Block Storage\\
					& Lustre \cite{Lustre}			& Centralised Object Storage\\
					& Ceph \cite{Weil2006} 			& Decentralised Object Storage\\
					& FusionFS \cite{Zhao2012} 		& ZHT, Data Access Paritions\\
Remote access DFS 	& Cloud Storage \cite{AmazonWebServices,Google} 	& Google Cloud Storage, S3, Azure Blob  \\
					& Container Storage \cite{Hindman2017} & DFS For Containers \cite{Quobyte}\\
Message Brokers 	& Apache Kafka \cite{JayKreps} 	& Log Structured Broker  \\
					& MQTT \cite{Locke}  			& Lightweight Pub/Sub Protocol\\
					& ZeroMQ \cite{Hintjens2013}  	& Lightweight Messaging Library \\
					& Edgware Fabric \cite{IBMEdgwareFabric} & IoT Service Bus (Discovery+Routing) \\
Gaian Databases 	& GaianDb \cite{Bent2008}  		& Self-organising Network\\
Autonomous/ML DB	& Pelaton \cite{Pavlo2017} 		& Classify, Forecast, Optimise Workload\\
					& Panoply  \cite{Panoply2017}	& ML Self-Optimisation\\
Federated SPARQL 	& FedX \cite{Schwarte2011}  		& Optimise Join Order (Heuristic)\\
					& SPLENDID   \cite{Gorlitz2011} & Optimise Join Order (Statistical)\\
					& LHD \cite{Wang2013lhd}, DARQ \cite{Quilitz2008} & Optimise Join Order (Statistical)\\
In-memory			& H-Store \cite{Kallman2008}  	& Partitions, Stored Procedures\\
					& MemSQL \cite{MemSQL} 			& Real-time Data Warehouse\\
					& Spark \cite{Zaharia2012} 		& In-memory DAG execution\\
MPP/Parallel Databases & Greenplum \cite{greenplum} & Master, Segment PostgresSQL  \\
					& Teradata Database \cite{Teradata} & Shared Nothing OLTP/OLAP\\
					& Volcano \cite{Graefe1994} 		& Exchange Meta-operator\\
Data Parallel		& Hadoop/MapReduce \cite{Dean2008a} & Big Data Programming Model\\
					& Dryad \cite{Isard2007}  		& Data Parallel App Runtime\\
Graph Parallel 		& GraphX \cite{Xin2013} 			& Resilient Distributed Graph Transform\\
					& GraphLab \cite{Low2012}  		& Asynchronous, Dynamic Computation\\
Cloud Compute 		& Virtualisation 				& EC2 \cite{AmazonWebServices}, Compute Engine \cite{Google} \\
					& Serverless 					& Lambda \cite{AmazonWebServices},  Functions \cite{Google,Microsoft} \\
					& Container	\cite{Casalicchio2016} & Portability, Overhead, Orchestration \\
Edge/Fog Compute 	& ANGELS \cite{Mukherjee2014} & Partition Data, Schedule Fog Jobs\\
					& Eywa \cite{Siow2017} 			& Distributed Stream Processing\\
 					& Cloudlets \cite{Satyanarayanan2009} & Proximity, Virtualisation\\
					& Cisco IOx \cite{CiscoIox} 		& Fog Director, App Host/Manage\\
\bottomrule
\end{tabular}
\end{center}
\centering
\footnotesize{\textbf{Legend:} DB=Database, DFS=Distributed File System, ML=Machine Learning, MPP=Massively Parallel Processing, OLAP/OLTP=Online Analytical/Transaction Processing, ZHT=Zero-hop Distributed Hash Table}
\end{minipage}
\end{table}%

\subsection{Storage Technologies}

Storage file systems need to cope with the huge amount of data from the IoT and work on `exascale' filesystems by Raicu \emph{et~al.} \cite{Raicu2011} look to address issues of scalability to millions of nodes and billions of concurrent input/output requests. The idea is to combine advances in non-volatile storage with those of distributed file systems. These include the management of distributed metadata, partitioning and knowledge of data access patterns to maximise data locality, resilience and high availability, data indexing and cooperative caching. An implementation exists in the form of FusionFS \cite{Zhao2012} which implements a zero-hop distributed hash-table (ZHT) for metadata management.

Similar decentralised distributed file systems (DFS) like Ceph \cite{Weil2006} and GlusterFS also manage metadata in a distributed way while other DFS like HDFS, which is part of the Hadoop ecosystem from Section \ref{subsec:archi}, iRODS and Lustre \cite{Lustre} are centralised with a single or replicated metadata servers. This group of DFS are classified as locally managed DFS and are compared in a survey by Depardon \emph{et~al.} \cite{Depardon2013}. Another group of DFS are remote access DFS like cloud storage from Google Cloud Storage \cite{Google}, S3 \cite{AmazonWebServices} and Azure Blob \cite{Microsoft}. Another interesting dimension to remote access DFS is the emerging Container Storage Interface (CSI) specification \cite{Hindman2017} for provisioning and managing storage, including cloud DFS like Quobyte \cite{Quobyte}, from container applications.

Bent \emph{et~al.} \cite{Bent2008} have designed a distributed, federated database architecture, Gaian Databases, that uses biologically-inspired, self-organising principles to organise a network of heterogenous relational or flat file databases and enable queries across them through query flooding. The work has become part of IBM's Smarter Planet \cite{IBMSmarterPlanet} project - an IoT-related vision of the planet that together with Edgware Fabric \cite{IBMEdgwareFabric} form a middleware layer for analytics and intelligence. The advantage of Gaian Databases is that through minimising network diameter and maximising connections to fit nodes, analytical queries on distributed data can be performed quickly and reliably.

Linked Data was seen as an approach to the aggregation step previously and work to access Linked Data across distributed sources has led to the area of federated querying. FedX \cite{Schwarte2011}, SPLENDID \cite{Gorlitz2011}, LHD \cite{Wang2013lhd} and DARQ \cite{Quilitz2008} are all engines that optimise federated query performance. They achieve improved performance by optimising the join order in queries. FedX takes a heuristic approach while the other engines take statistical approaches. Saleem \emph{et~al.} \cite{Saleem2014} and Hartig \cite{Hartig2013} review and evaluate the systems. More specific performance bottlenecks like data distribution \cite{Rakhmawati2012} and other challenges \cite{Rakhmawati2013} for federated engines have still to be addressed though. 

A message broker is an intermediary that routes a message from publishers to subscribers. A message broker can serve as a storage and interoperability technology in distributed systems as it can provide a formal message protocol for publishing and subscribing, reliable storage and guaranteed message delivery. Log-structured storage has been utilised for high throughput distributed message brokers like Kafka \cite{JayKreps} or the scalable data middleware for smart grids described by Yin \emph{et~al.} \cite{Yin2011}. MQTT \cite{Locke}, a lightweight publish-subscribe protocol for the IoT, ZeroMQ \cite{Hintjens2013}, a messaging protocol library and Edgware Fabric \cite{IBMEdgwareFabric}, an IoT service bus, are other examples of technologies used in distributed message broker systems.

Autonomous or self-driving database management systems like Pelaton \cite{Pavlo2017} integrate artificial intelligence components to automatically classify and forecast workloads so that the database can optimise physical storage, data location and partitioning in distributed or cloud-based environments and runtime resources, configuration and query cost models. Panoply \cite{Panoply2017} is a similar machine learning optimised autonomous data warehouse, which additionally allows the ``self-preparation'' (automated transformation and integration) of ingested semi-structured data.

Massive Parallel Processing (MPP) databases are build on top of shared-nothing MPP grids where data is sharded between nodes and nodes processes computations, queries to retrieve and process data, in parallel. Greenplum \cite{greenplum}, which uses a master-segment approach with each segment a PostgresSQL database, and Teradata \cite{Teradata} are examples of MPP databases. Volcano \cite{Graefe1994} was an early system that presented research on parallelising query operators through the exchange of meta-operators.
%

\subsection{Compute Technologies and IoT Analytics Applications}
\label{subsec:compute_tech}

The elasticity of resources on the cloud is often considered an advantage for deploying horizontally-scalable parallel processing paradigms that work with big IoT data. Compute on the cloud can be divided into virtualisation, serverless computing and container technologies and orchestration. Major vendors like the Google Cloud Platform \cite{Google}, Amazon Web Services \cite{AmazonWebServices} and Microsoft Azure \cite{Microsoft} each have options for virtualisation, Compute Engine, EC2 and Azure Virtual Machines respectively, which allow a full server to be provisioned for compute and storage tasks. Each also supports the serverless execution of compute functions through Cloud Functions, Lambda and Azure Functions respectively. Finally, container technologies \cite{Casalicchio2016} are becoming increasingly popular as they increase application portability and reduce dependencies, have lower overhead and faster launch times than virtual machines and the orchestration of containers allows the efficient provisioning, deployment and management of distributed compute clusters. Kubernetes, Docker Compose \cite{Tosatto2015} and Mesosphere \cite{Mesosphere2017} are such container orchestration technologies.

Various IoT infrastructures and deployments have implemented distributed cloud-based compute. Xu \emph{et~al.} \cite{Xu2014a} have developed a cloud-based time-series analytics platform for the IoT that stores and indexes time series data, analyses and mines for patterns and allows searching on patterns and abnormal pattern discovery. Indexes specifically optimised for time-series data help achieve real time analytics at a lower latency (at the cost of increased storage space). Ding \emph{et~al.} \cite{Ding2013}, propose a means of doing statistical analysis on the cloud. Spatial aggregation of the area in a city where the pollution level is above a certain threshold or parameter aggregation to calculate the average pollution level at a certain time in a city are examples. The novel part of this approach is that analytics is implemented within the database kernel itself, improving performance by reducing the transfer of data (to the master node for processing).  

Nastic \emph{et~al.} \cite{Nastic2013} have designed a high level programming model abstraction for the IoT running on the cloud. In the model, there are abstractions called Intents and Intent Scopes which describe a task and a group of `things' respectively, from underlying distributed and heterogenous sources that share a common context. By coupling Intents and Intent Scopes with analytics operators, complex IoT applications can be designed, optimised on a distributed compute system, and run on the cloud. Guazzelli \emph{et~al.} \cite{Guazzelli2009} make use of the Predictive Model Markup Language (PMML) \cite{PMML2015}, an XML based markup language to describe data mining models, to run analytics on the cloud. Web service calls can be made to instances on the cloud, submitting markup that then execute tasks like running regression models, clustering, learning based on artificial neural networks (ANN), decision trees, support vector machines or mining association rules. 

Next, we briefly summarise specific distributed compute technologies from the small programming constructs and components used to build distributed compute to the large big data systems made from these components which we divide into: in-memory and stream systems, parallel programming models, graph parallel models and edge/fog computing systems.

A means that nodes in a distributed compute system can communicate is through message passing. A Remote Procedure Call (RPC) is a form of message passing and gRPC, a multiplexed, bi-directional streaming RPC protocol, and Thrift \cite{prunicki2009apache}, an asynchronous RPC system, are examples. The Actor Model \cite{Haller2012} is a message passing programming model supporting asynchronous communication in distributed compute systems the provides an abstraction enabling looser coupling among components, allows for behaviour reasoning, and provides a lightweight concurrency primitive across machines. Futures or Promises are another construct for asynchronous programming and are abstractions of values that will eventually become available. They are useful for message passing in distributed compute to reason about state changes when latency is a concern. 

Distributed in-memory databases like H-store \cite{Kallman2008} and MemSQL \cite{MemSQL} allow low latency stored procedures and interactive querying respectively to be done on scale-out transactional databases, hence, overcoming memory limitations by adding nodes. They are so fast that they can be used for real-time compute tasks rather than just storage. Spark \cite{Zaharia2012} is an in-memory data processing engine with two main abstractions, an immutable, read-only collection of objects within Resilient Distributed Datasets (RDD) (as opposed to fine-grained Distributed Shared memory (DSM)) and parallel operations represented as an acyclic data flow graph. SparkSQL includes an execution model that uses the Catalyst query optimiser \cite{Armbrust2015} for both rule-based and cost-based optimisation to form a Spark data flow graph. D-streams is the Spark streaming abstraction where a streaming computation is treated as series of deterministic batch computations on RDDs within small time intervals. This type of stream processing is called micro-batch processing while Complex Event Processing (CEP) \cite{Luckham2002} involves continuous operators on each tuple. Khare \emph{et~al.} \cite{Khare2015} show how continuous operators can work on publish-subscribe IoT sensor streams with a Functional Reactive Programming (FRP) language. The system was tested on sensor data of a football match to aggregate running data for each player and create descriptive analytics heat maps for players.

Data parallelism means that each node in a distributed compute system can perform independent calculations on a meaningful subset of data. MapReduce \cite{Dean2008a}, of which Hadoop MapReduce (Section \ref{subsec:archi}) is an implementation, and Dryad \cite{Isard2007} are both programming models for data parallel processing on big data. Hammond \emph{et~al.} \cite{Classification2013} deploy analytics in the cloud using Hadoop MapReduce. The analytics techniques include text classification using Naive Bayes, a top-K recommendation engine based on similarity and a Random Forests classifier to categorise data as part of Decision Support Systems. MapReduce, however, does not scale easily for iterative graph algorithms as each  iteration requires reading and writing results to disk. Graph Parallel abstractions like those in GraphX \cite{Xin2013}, for graph transformations, and GraphLab \cite{Low2012}, for asynchronous computation, support these.

Finally, Fog or Edge Computing technologies like ANGELS \cite{Mukherjee2014} and the scheduler designed by Dey \emph{et~al.} \cite{Dey2013} propose utilising the idle computing resources of edge devices like smartphones through a scheduler in cloud. The edge devices themselves keep track of their resource usage states, which are formed based on user behavioural patterns, and advertise free slot availability. The cloud servers receive analytics jobs and advertisements from edge devices and then schedule subtasks to these devices. Distributed stream processing within a fog computing network has also been implemented in the Eywa framework \cite{Siow2017} using inverse-publish-subscribe for the control plane and workload pushdown to fog nodes for projections in the data plane. Cloudlets \cite{Satyanarayanan2009} allow a mobile user to instantiate virtualised compute tasks on physically proximate cloudlet hardware. Cisco IOx \cite{CiscoIox} is another platform that consists of a fog director, application host and management components, allowing fog computing tasks to be virtualised and executed on fog nodes.

\subsection{Levels of Distribution of Storage and Compute}

\begin{figure}
\centering
\includegraphics[width=0.7\textwidth]{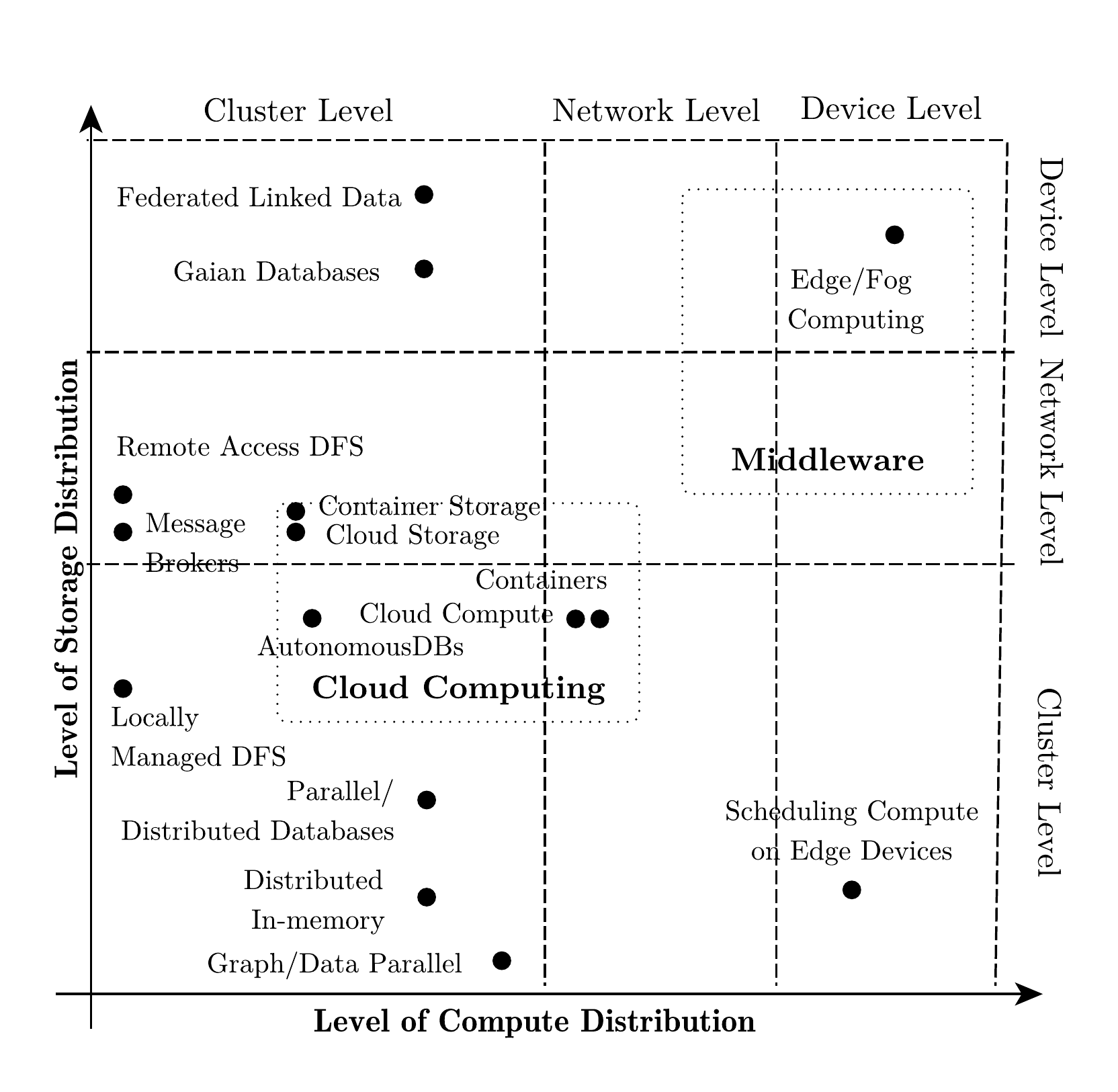}
\caption{Technology and their Levels of Distribution for Storage and Compute}
\label{fig:distribution}
\end{figure}

The Internet of Things, as defined in Section \ref{sec:iot}, comprises smart and interconnected physical objects with varying storage and compute capabilities. Analytics processing can be done at different distribution levels depending on how far data from physical objects can and should travel and on the storage and compute capabilities at each of:
\begin{enumerate}
\item the device level, where devices act not just as data producers but as participants of the storage and compute process,
\item the network level, involving remote connections to fog computing nodes, hubs, base stations, gateways, routers and servers and
\item the cluster level, within a group of interconnected servers.
\end{enumerate}
Enabling infrastructure and technologies are observed to address each of these levels of distribution of compute and storage to a different degree. A classification of the surveyed IoT enabling technologies is proposed in Fig. \ref{fig:distribution} along the axes of storage and compute distribution.

At the cluster level, we see storage systems that are distributed within locally managed clusters. Usually these clusters are located within data centres and connected by top-of-the-rack switches in a hierarchical fashion (intra and inter rack). Locally managed distributed file systems are an example. In-memory systems distribute both processing and storage, usually by partitioning the data onto nodes, in a centrally managed cluster and running processing on each node that corresponds with the data on that node. Similarly, Massive Parallel Processing Databases, Data Warehouses and Parallel/Distributed Databases are examples of systems with distributed storage and compute on each node. Examples of compute within clusters of distributed servers include Parallel Processing frameworks like MapReduce \cite{Dean2008a}.

Cloud Computing is usually divided into private, public or hybrid clouds. Private clouds share similarities and types of distributed storage and compute with those previously mentioned at cluster level. Public clouds are remotely managed and hence belong to the network level of distribution of which Cloud Storage and Cloud Compute Engines serve storage and compute tasks respectively. Hybrid clouds bridge both public and private clouds. Similar to cloud storage are remote access distributed file systems. Message brokers, message queues and log-based systems are some other examples of network level, possibly remote access storage systems.

At the device level of storage, we have technologies like federated Linked Data endpoints and Gaian databases where data can reside on their respective devices but be accessed by other clients. On the device level for compute, scheduling of compute tasks on fog or  edge devices, is an example. Finally, both compute and storage distribution from the network to device level is present in Edge and Fog computing and middleware is usually used to connect such edge systems together.

Table \ref{table:products} summarises the surveyed literature on distributed storage and compute technologies to provide a point of reference for researchers on current state-of-the-art implementations.

This review of enabling infrastructure and technologies at each part of the data flow process, classification of the storage and compute distribution and examples of distributed storage and compute technologies form a basis for a direction of future work towards tackling the challenges big data analytics on the IoT.

\section{Research Challenges}
\label{sec:challenges}

\begin{figure}[!h]
\centering
\includegraphics[width=0.5\textwidth]{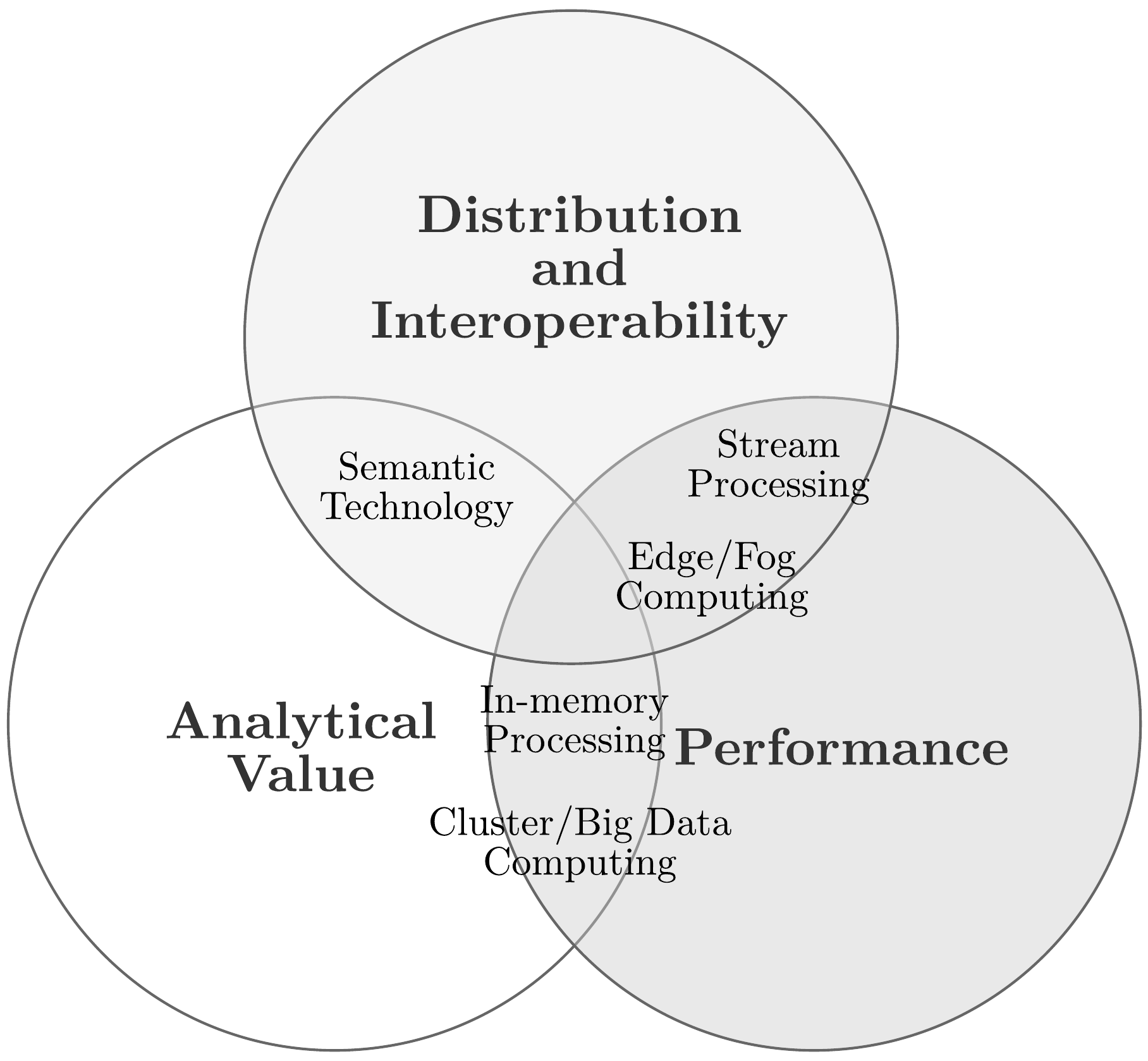}
\caption{Tradeoffs in Designing for Analytics on the IoT}
\label{fig:tradeoffs}
\end{figure}

As we have seen in our study of enabling infrastructure and present analytical applications in the IoT, there are still some challenges that we face in aligning the vision of the IoT with that of analytics. In particular, we argue that infrastructure for analytics in the IoT faces a tradeoff between:
\begin{enumerate}
\item Distribution \& Interoperability, complicated by big data \emph{variety}, 
\item Performance, complicated by the \emph{volume and velocity} of the big data problem,
\item and Analytical Value, which deals with how high the output of analytics applications is on the knowledge hierarchy from Fig. \ref{fig:knowledgehierachy}.
\end{enumerate}
Fig. \ref{fig:tradeoffs} depicts the tradeoffs that IoT infrastructure for analytics applications face in terms of these three challenges. For example, the semantic technology community argues for its utility in the IoT \cite{Barnaghi2012} to encourage semantic interoperability, while semantic ontologies provide analytical value and federation supports diverse, heterogeneous distributed sources. Performance of such systems though are still questionable \cite{Saleem2014, Rakhmawati2013}. Edge and fog computing is also an emerging area of distributed technologies that promises advantages in latency for real-time processing of streams and efficiency due to its proximity sources \cite{Chiang2017}. However, cloud-based clusters and fast distributed OLTP in-memory processing still offer greater analytical value combining big data sans the advantages of IoT distribution and interoperability.

Variety has been a less researched aspect of the big data problem but is apparent in the IoT paradigm. Heterogenous data sources in the IoT combined with the need for analytics to also involve a wide range of multi-modal data sources like social media, Linked Data, image and video data, satellite and geospatial data, voice data, etc. makes the variety problem highly analogous with the richness of insights and knowledge that can be derived in analytics applications. Predictive analytics can be made more accurate through corroboration of independent data sources and prescription can be optimised with more and varying knowledge inputs. Solving the variety problem can be seen as an opportunity to enhance the value of current IoT applications.

Performance and scalability questions still exist in current systems because of the scale of the IoT. This is not only about scaling the storage of data or of the communications layer but also the scaling of infrastructure to do analytics processing. We see distributed analytics as a plausible means of handling IoT scale-data (which is predicted to be larger and richer than web scale data) and there is potential for more work in this area.

\section{Conclusions}
\label{sec:conclusion}
The Internet of Things (IoT) has huge potential to provide advanced services and applications across many domains and the momentum that it has generated, together with its broad visions, make it an ideal frontier for pushing technological innovation. We have shown that analytics plays a role in many applications, across many domains, designed for the IoT and will be even more important in the future as the enabling infrastructure develops and scales and the deployment of devices becomes truly ubiquitous. We have applied a systematic review of analytics applications in the IoT to the task of understanding analytics as it develops. This results in a layered taxonomy that defines and categorises analytics by their capabilities and application potential for research and application roadmaps. We then review the enabling infrastructure and discuss the technologies from different stages in the data flow for analytics. Finally, we look at some tradeoffs for analytics in the IoT that can shape research direction going forward.

\bibliographystyle{ACM-Reference-Format}
\bibliography{IntenetofThings.bib}

\end{document}